\documentclass[twocolumn,preprintnumbers,pre]{revtex4-2}
\usepackage{graphicx, subfigure}
\usepackage{amssymb, amsmath,amssymb,amsfonts}
\usepackage{amsthm,mathrsfs,amsopn}
\usepackage{array}  
\usepackage{dcolumn}
\usepackage{bm}
\usepackage{soul}
\usepackage{enumerate}
\usepackage{color}
\usepackage[utf8]{inputenc}
\usepackage{mathtools}
\theoremstyle{plain} 

\usepackage{xr}
\begin{document}

\title{Studying speed-accuracy trade-offs in best-of-n collective decision-making\\through heterogeneous mean-field modeling}

\author{Andreagiovanni Reina}
\affiliation{Institute for Interdisciplinary Studies on Artificial Intelligence (IRIDIA), Universit\'{e} Libre de Bruxelles, Brussels, Belgium;\\
Centre for the Advanced Study of Collective Behaviour, Universit\"{a}t Konstanz, Konstanz, Germany;\\
Department of Collective Behavior, Max Planck Institute of Animal Behavior, Konstanz, Germany
\\
andreagiovanni.reina@gmail.com}
\author{Thierry Njougouo}
\author{Elio Tuci}
\affiliation{Faculty of Computer Science and Namur Institute for Complex Systems, naXys\\
Universit\'{e} de Namur, Rue Grandgagnage 21, B5000 Namur, Belgium}
\author{Timoteo Carletti}
\affiliation{Department of Mathematics and Namur Institute for Complex Systems, naXys\\
Universit\'{e} de Namur, Rue Graf\'e 2, B5000 Namur, Belgium\\
timoteo.carletti@unamur.be}

\begin{abstract}
To succeed in their objectives, groups of individuals must be able to make quick and accurate collective decisions on the best option among a set of alternatives with different qualities. Group-living animals aim to do that all the time. Plants and fungi are thought to do so too. Swarms of autonomous robots can also be programmed to make best-of-n decisions for solving tasks collaboratively. Ultimately, humans critically need it and so many times they should be better at it! Thanks to their mathematical tractability, simple models like the voter model and the local majority rule model have proven useful to describe the dynamics of such collective decision-making processes. To reach a consensus, individuals change their opinion by interacting with neighbors in their social network. At least among animals and robots, options with a better quality are exchanged more often and therefore spread faster than lower-quality options, leading to the collective selection of the best option. With our work, we study the impact of individuals making errors in pooling others' opinions caused, for example, by the need to reduce the cognitive load. Our analysis is grounded on the introduction of a model that generalizes the two existing models (local majority rule and voter model), showing a speed-accuracy trade-off regulated by the cognitive effort of individuals. We also investigate the impact of the interaction network topology on the collective dynamics. To do so, we extend our model and, by using the heterogeneous mean-field approach, we show the presence of another speed-accuracy trade-off regulated by network connectivity. An interesting result is that reduced network connectivity corresponds to an increase in collective decision accuracy. 
\end{abstract}

\maketitle
\section{Introduction}
\label{sec:intro}
Reaching a consensus in a group of individuals without any central authority or coordinator requires individuals to exchange opinions and combine conflicting information received from peers. Studying the situation in which the group must agree on the best among a set of options---the so-called best-of-n problem---is interesting because it helps us to both understand biological processes and design the robotics systems of our future~\cite{Reina:SwInt:2021,Valentini2017}. Social insects are an example of collectives which need to solve the best-of-n problem when selecting the site where to nidificate~\cite{Britton2002,Seeley2012,Reina:PRE:2017}. While each insect makes an inaccurate estimate of the quality of each site, the colony is able to filter noise and reach a consensus on the best alternative \cite{Passino2007}. Similarly, other more complex animals make collective decisions on when and in which direction to flee danger, or the location where to forage~\cite{Davis2022,Conradt2009,Ward2008}. Collective agreement is achieved by individuals sharing their opinion with others (voting) and, in turn, adopting the opinion expressed by others' votes. These simple voting rules employed by animals are a useful source of inspiration to design algorithms for robot swarms, which make best-of-n decisions, for example, on the shortest path to navigate \cite{Scheidler2016,Reina:SwInt:2015} or the most important location for their operations \cite{Reina:DARS:2016,Valentini2016}.

The group is able to select the best alternative because each individual shares her opinion as frequently as the estimated quality, that is, better alternatives are shared (voted) more often~\cite{Marshall2009,Valentini2014}. Despite the individual estimates being incorrect, most of the time the group reaches a consensus for the option that, on average, is estimated to be of higher quality. Depending on the effort individuals make in acquiring, processing, and sharing information in their social network, the collective dynamics change, e.g., in the group accuracy or the decision speed. While there are several studies analyzing voting models in decentralized networks \cite{Sood2005,Redner2018,Moretti2013,moinet2018generalized}, there is no explicit connection between the individual cognitive requirements, the social network, and the collective decision-making performance.

Here, we build a model that explicitly considers the cognitive effort that individuals make in acquiring and processing their neighbor's vote. The more cognitive effort individuals put in, the better they pool social information (making smaller pooling errors). We investigate how speed and accuracy change in collective decisions both as a function of the individual cognitive load and the interaction network. Our analysis reveals that an increased cognitive effort leads to quicker and more democratic collective decisions, but which are not necessarily accurate. Counter-intuitively, the highest levels of collective decision accuracy can be achieved with moderate levels of cognitive efforts. This however comes at the expense of a longer deliberation time. Thus, our model enriches our understanding of the classical speed-accuracy trade-off in decision-making \cite{Passino2006,Marshall2006,Valentini2016,Talamali:ICRA:2019,Daniels2021} by describing it through the lens of the individual cognitive load.

Additionally, the network analysis we performed reveals that groups that are sparsely connected can obtain higher collective accuracy than when they are highly connected. Recent previous research has shown that in a number of conditions, having reduced connectivity between the group members can improve collective performance in terms of coordination, accuracy, or response speed \cite{Sood2008,Sosna2019,Mateo2019,Rahmani2020,Talamali:SciRobot:2021,Hiraga2023}. More precisely, fish adaptively change their interaction network when exposed to a threat in order to maximize information transfer in the fish school \cite{Sosna2019}. Robots that can only run simplistic algorithms can also exploit the advantages of sparse connections to improve swarm accuracy \cite{Talamali:SciRobot:2021,Aust:ANTS:2022}. While it is commonplace to assume that higher connectivity can improve opinion sharing and thus lead to better coordination, these recent results show in which conditions limited connectivity can lead to improved collective dynamics. Our analysis uses the Heterogeneous Mean-Field theory (HMF)~\cite{PSV2001,CPSV2007,pastor2015epidemic,costa2022heterogeneous} to show that both network connectivity and individual cognitive load can be control parameters to regulate the speed-accuracy trade-off of group decision-making. 

\section{The model}
\label{sec:themodel}
Let us consider a population composed of $N$ agents making a binary collective decision between two alternative options, say $A$ and $B$. Each option is characterized by a quality, $Q_A$ and $Q_B$ (for option $A$ and $B$, respectively); without lack of generality we hereby assume $Q_A>Q_B>0$ and we will define the quality ratio $Q=Q_B/Q_A \in (0,1)$. Each agent, at a given time $t$, has an opinion in favor of either option, $A$ or $B$. Throughout the collective decision-making process, agents interact with each other and change their opinions depending on the votes expressed by their neighbors. Each agent votes with a frequency linearly proportional to the estimated quality of each option, thus $Q_A$ and $Q_B$ for options $A$ and $B$, respectively. Therefore, through mean-field approximation, we model the change of agents' opinions as a function of the number of agents with opinion $A$ and $B$, denoted by $n_A(t)$ and $n_B(t)$, respectively, weighted by the respective option's quality and normalized by the group size $N$. Such weighted proportions, $n_A^\#$ and $n_B^\#$, represent the mean-field approximation of the votes expressed by the agents in favor of option $A$ and $B$, respectively, and correspond to
\begin{equation}
\begin{split}
\label{eq:defnBdies}
n_A^\# = \frac{Q_A n_A/N}{Q_A n_A/N+Q_B n_B/N} & \qquad \text{and} \\
n_B^\# = \frac{Q_B n_B/N}{Q_A n_A/N+Q_B n_B/N}\,.
\end{split}
\end{equation}

Aiming at reaching a group consensus, agents follow a conformist rule where they align their opinion with the most voted opinion by their neighbors. However, when agents put a limited effort into acquiring and processing others' votes because, for example, they need to reduce their cognitive load, they may wrongly compute what is the predominant opinion in their neighborhood, making what we call a pooling error. Such pooling errors can be caused by agents that, for example, subsample their neighborhood (i.e., they do not record the votes from all their neighbors but only from a subset of them), or occasionally record the incorrect opinion of some of their neighbors. Therefore, the pooling error $\alpha$ is inversely linked to the effort that the agent invests into pooling social information. When the agents put maximum effort (corresponding to maximum cognitive load), we can assume they do not make any pooling error, $\alpha=0$. This case corresponds to the weighted local majority rule model where each agent collects all the votes of its neighbors, group them by opinion, and adopts the opinion voted by the majority. Instead, when agents only sample a single vote from a randomly-selected neighbor and adopt her opinion, they commit moderate levels of pooling error (in our model, $\alpha = 1$). This case corresponds to the weighted voter model, where the probability that an agent committed to $A$ changes her opinion to $B$ is equal to the weighted proportion $n_B^\#$. The extreme case of maximum pooling error $\alpha \gg 1$ corresponds to agents changing their opinion totally ignoring others' votes. Despite the high error, agents experience very low cognitive load as they do not make any effort to coordinate with the others. 

Previous work has investigated opinion dynamics in populations of agents that update their opinion through either the voter model~\cite{Clifford1973,Holley1975,Jhawar2020,FernandezGracia2014,Zillio2005,Redner2018,de2020emergence} (later extended to the weighted voter model~\cite{Valentini2014}) and the local majority rule model~\cite{Galam1986,Krapivsky2003,Kao2014,krapivsky2021divergence} (later extended to the weighted local majority rule model~\cite{Scheidler2011,Montes2010,Valentini2015aamas}). We build a model that generalizes the two existing (weighted) models and can also interpolate, in a continuous way, the cognitive load level in the form of pooling error among the two models and beyond. In our model, agents change their opinion with probability
\begin{equation}
\label{eq:Px}
 P_\alpha(x) = 
            \begin{dcases}
             \frac{1}{2}-\frac{1}{2}\left(1-2x\right)^{\alpha} & \text{if $ 0\leq x \leq \frac{1}{2}$}\\
              \frac{1}{2}+\frac{1}{2}\left(2x-1\right)^{\alpha}  & \text{if $ \frac{1}{2} < x \leq 1$}\,,
            \end{dcases}\, 
\end{equation}
where $\alpha \geq 0$ is the pooling error and $x \in [0,1]$ is the weighted proportion of agents with a different opinion. Therefore, an agent with an opinion in favor of option $A$ (resp. $B$) will change her opinion to $B$ (resp. $A$) with probability $P_\alpha(n_B^\#)$ (resp. $P_\alpha(n_A^\#)$).
Note that the assumption of fixed population, $n_A(t)+n_B(t)=N$, implies $n^\#_A(t)+n^\#_B(t)=1$ and therefore, as a consequence of the functional form of Eq.\,\eqref{eq:Px}, we have 
\begin{equation}
\label{eq:PaPB}
P(n_A^\#)+P(n_B^\#)=1\,.
\end{equation}
%
\begin{figure}[t!]
\centering
\includegraphics[width=0.4 \textwidth]{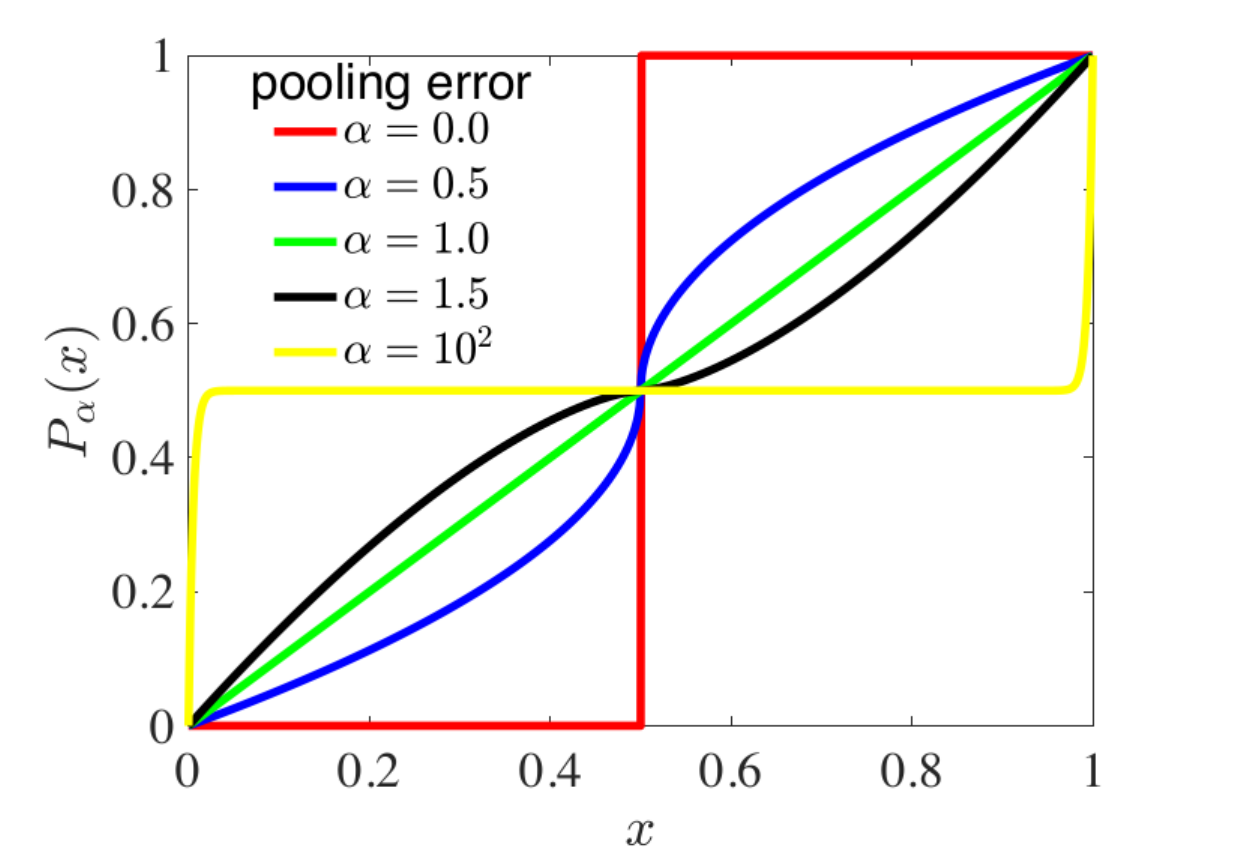}
\caption{The probability $P_\alpha(x)$ given by Eq.\,\eqref{eq:Px} for representative values of the pooling error $\alpha$, which is a parameter inversely proportional to agents' cognitive load. For $\alpha=0$, our model corresponds to the (weighted) local majority rule model~\cite{Krapivsky2003,Valentini2015aamas}, for $\alpha=1$ our model corresponds to the (weighted) voter model~\cite{Clifford1973,Valentini2014}, for $0<\alpha<1$ our model interpolates between the two, and for $\alpha>1$ the agents change their opinion with little attention to others' opinion.}
\label{fig:Px}
\end{figure}

Fig.~\ref{fig:Px} shows a graphical representation of $P_\alpha(x)$ of Eq.\,\eqref{eq:Px} and let us appreciate that intermediate values of the pooling error $\alpha$ allow interpolating between the two models. Indeed, values of $0 < \alpha < 1$ represent cases in which an agent makes a higher effort than sampling a random individual (as she does in the voter model), still the probability of changing opinion in favor of the most-voted option is lower than the `perfect' case (zero error) of the local majority rule model. These intermediate values represent conditions in which the agent samples only a subset of the population or approximately and imprecisely integrates others' votes. Values of $\alpha>1$ further reduce the cognitive effort that the agent puts into taking into consideration others' opinions. As the value of the pooling error $\alpha$ increases, the probability $P_\alpha(x)$ gradually becomes more and more independent of the actual votes. For $\alpha \gg 1$, the probability of changing opinion approximates the flat line $P_\alpha(x)=0.5$, that is, the agents make maximum levels of pooling error by randomly changing their opinion regardless of the opinions expressed by the others.

We first consider a well-mixed population, that is, agents are the nodes of a complete network and therefore every agent can directly exchange votes with all the other agents. Despite being an idealized case, it allows us to build a deep analytical understanding of our model in Section~\ref{sec:mf}, preliminary to the study of a population interacting on a heterogeneous network in Section~\ref{sec:HMF}.

\section{Mean-field analysis}
\label{sec:mf}
Let us introduce the proportion of agents with opinion $A$ (resp. $B$), $a(t)=n_A(t)/N$ (resp. $b(t)=n_B(t)/N$), hence $a(t)+b(t)=1$. The proportion of agents with opinion $A$ increases when agents with opinion $B$ change their minds and adopt opinion $A$, or decreases when agents with opinion $A$ adopt opinion $B$. As illustrated in Appendix~\ref{sec:compeltegraph}, exploiting the well-mixed hypothesis, we obtain the time evolution of $a(t)$ in the form of the following ordinary differential equation
\begin{equation}
\label{eq:odea2}
\frac{da}{dt} = - a + P_\alpha\left( \frac{a}{a(1-Q)+Q}\right) =:f_\alpha(a)\, .
\end{equation}
Because the population size is finite and fixed ($a+b=1$), Eq.\,\eqref{eq:odea2} is sufficient to fully determine the temporal dynamics of the system, without the need to explicitly define another equation ruling the evolution of the proportion of agents with opinion $B$.

We analyze the long-term dynamics of the system by finding the equilibria of Eq.\,\eqref{eq:odea2} and computing their stability as a function of the model parameters. The equilibria are found at values of $a$ that satisfy $f_\alpha(a)=0$. We find that $\hat{a}^*=1$ and $\check{a}^*=0$ are always two zeros of $f_\alpha(a)$. When these equilibria are stable, they correspond to a consensus decision for either alternative: for $\hat{a}^*=1$,  all agents eventually have opinion $A$, that is, the population has selected the best option (because $Q_A>Q_B$), or, when $\check{a}^*=0$, all agents eventually have opinion $B$ and therefore the population has made a collective mistake by selecting the option with the inferior quality. For a range of values of $\alpha$ and $Q$, a third equilibrium $\tilde{a}^* \in (0,1)$ may exist and it corresponds to a polarized population, in which agents with opinion $A$ and $B$ coexist. In this case, there is not a consensus decision but the population is in a decision deadlock. Recall that $Q<1$; therefore, we can prove (as detailed in Appendix~\ref{sec:compeltegraph}) that 
\begin{enumerate}[i)]
\item if $Q > \alpha $, both $\check{a}^*=0$ and $\hat{a}^*=1$ are {\em stable} equilibria, and a third equilibrium $0<\tilde{a}^*<1$ exists and is {\em unstable};
\item if $Q<\alpha < 1/Q$, then $\check{a}^*=0$ is {\em unstable} while $\hat{a}^*=1$ is {\em stable}, the third equilibrium $0<\tilde{a}^*<1$ does not exist;
\item if $1/Q<\alpha$, both $\check{a}^*=0$ and $\hat{a}^*=1$ are {\em unstable} equilibria, and a third equilibrium $0<\tilde{a}^*<1$ exists and is {\em stable}.
\end{enumerate}
Note that when the third equilibrium $\tilde{a}^*$ exists and it is unstable -- i.e., the case (i) above -- the fate of the system depends on the initial conditions. Stated differently, the position of the third equilibrium $a^*$ splits the interval $[0,1]$ into two parts $[0,\tilde{a}^*)$ and $(\tilde{a}^*,1]$, and if the initial conditions are such that $a(0)\in [0,\tilde{a}^*)$, then $a(t)\rightarrow 0$ (and thus $b(t)\rightarrow 1$), while if $a(0)\in(\tilde{a}^*,1]$, then $a(t)\rightarrow 1$ (and thus $b(t)\rightarrow 0$). If there are only two equilibria -- i.e., the case (ii) above -- the system converges to $\hat{a}^*=1$ for any initial conditions (which correspond to the accurate collective decision, being $Q_A>Q_B$).

\begin{figure}[t!]
    \centering
    \includegraphics[width=0.4\textwidth]{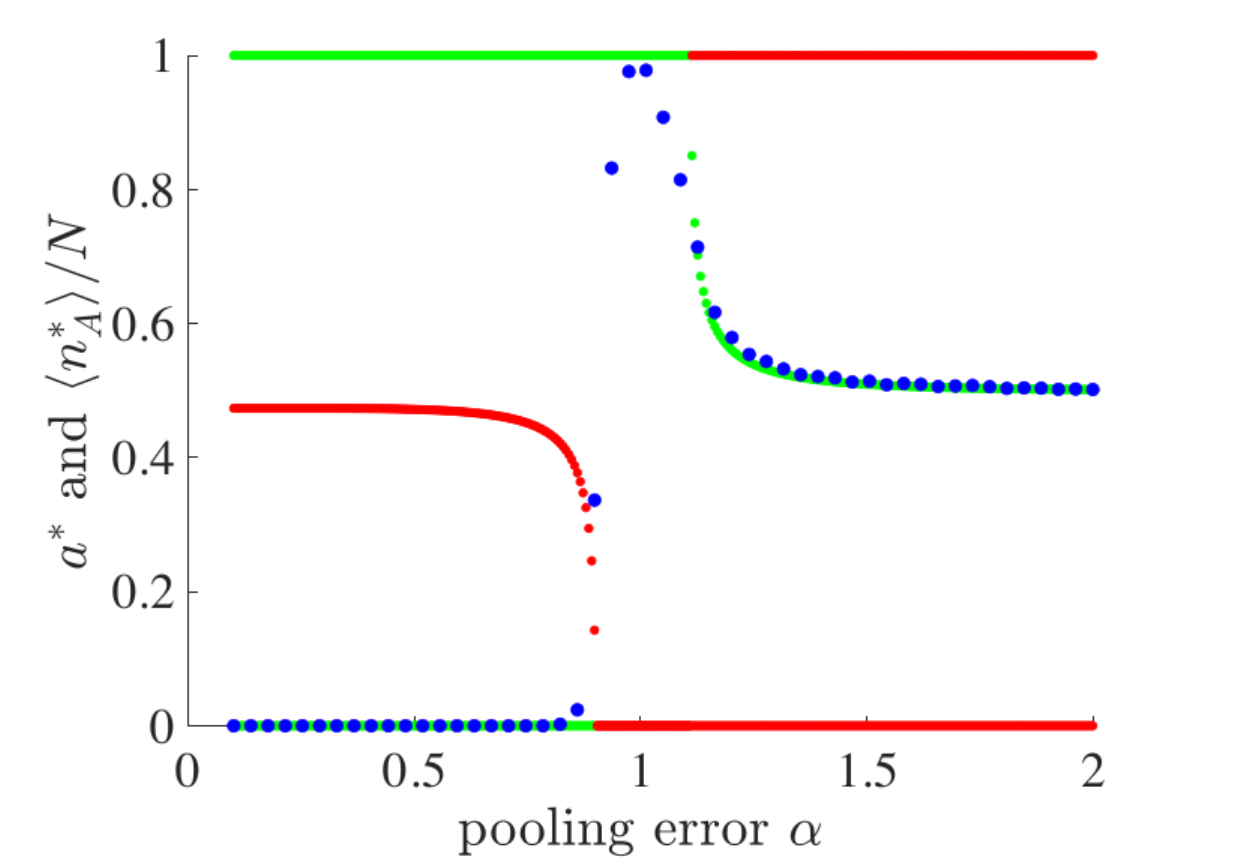}
    \caption{Bifurcation diagrams showing the equilibria of the mean-field model of Eq.\,\eqref{eq:odea2} as a function of $\alpha \in [0,2]$ for $Q_A=1$ and $Q_B=0.9$. Green dots show stable equilibria, red dots show unstable equilibria, and blue dots show the average asymptotic values of $\langle n_A\rangle/N$ obtained with stochastic numerical simulations of a population of $N=500$ agents interacting on a complete graph and initialized at $n_A(0)=100$. The match between mean-field model predictions and simulations is good.}
    \label{fig:figEquil}
\end{figure}

In Fig.~\ref{fig:figEquil}, we show the bifurcation diagram of the mean-field model of Eq.\,\eqref{eq:odea2} for $Q_A=1$ and $Q_B=0.9$, as a function of $\alpha$. We report the equilibria and their stability (in green when stable and in red when unstable) for values of $\alpha$ in the range $[0,2]$ (x-axis). Note that $\check{a}^*=0$ (corresponding to a population fully committed to the inferior option $B$) is a stable equilibrium for $\alpha < Q$ and unstable otherwise. Similarly, $\hat{a}^*=1$ (corresponding to a population fully committed to the best option $A$) is a stable equilibrium for $\alpha < 1/Q$ (that is $1/0.9\approx1.11$ for the used values of $Q_A$ and $Q_B$) and unstable in the remaining range of $\alpha$. The third equilibrium $0<\tilde{a}^*<1$, associated with a population where opinions $A$ and $B$ coexist, is unstable for $\alpha < Q$ (red branch), stable for $\alpha > 1/Q$ (green branch), and does not exist for $Q < \alpha < 1/Q$. 

In Fig.~\ref{fig:figEquil}, we also overlay to the analytical bifurcation diagram the results obtained from the numerical stochastic simulation of a system defined on a complete graph made of $N=500$ agents, using the same values for the opinion quality, $Q_A=1$ and $Q_B=0.9$. For each value of $\alpha \in [0,2]$, we numerically integrate the system for $50\,000$ timesteps and report the average of the final proportion of agents with opinion $A$ (i.e., $\langle n_A\rangle/N$) over 21 runs.
When $\alpha < Q_B$, the mean-field theory predicts that the system converges to $\check{a}^*=0$ if the initial condition is below the value of the third unstable equilibrium $\tilde{a}^*$. In agreement with the theoretical predictions, our simulations, which have been initialized with $n_A(0)=100$ (i.e., $a=0.2$), reached the final system state $n_A=0$ for $\alpha < Q$ (blue dots in Fig.\,\ref{fig:figEquil}). When only two equilibria exist, the mean-field analysis predicts convergence to $\hat{a}^*=1$ for any initial conditions, and indeed the blue points (up to finite size effects) converge to $n_A=N$. Finally for large $\alpha$, when the third equilibrium is stable, according to the mean-field analysis, the system should oscillate about this third equilibrium; the numerical results agree well with the analytical predictions, with the blue dots aligned with the green curve.

\begin{figure*}[htp!]
    \centering
        \includegraphics[width=\textwidth]{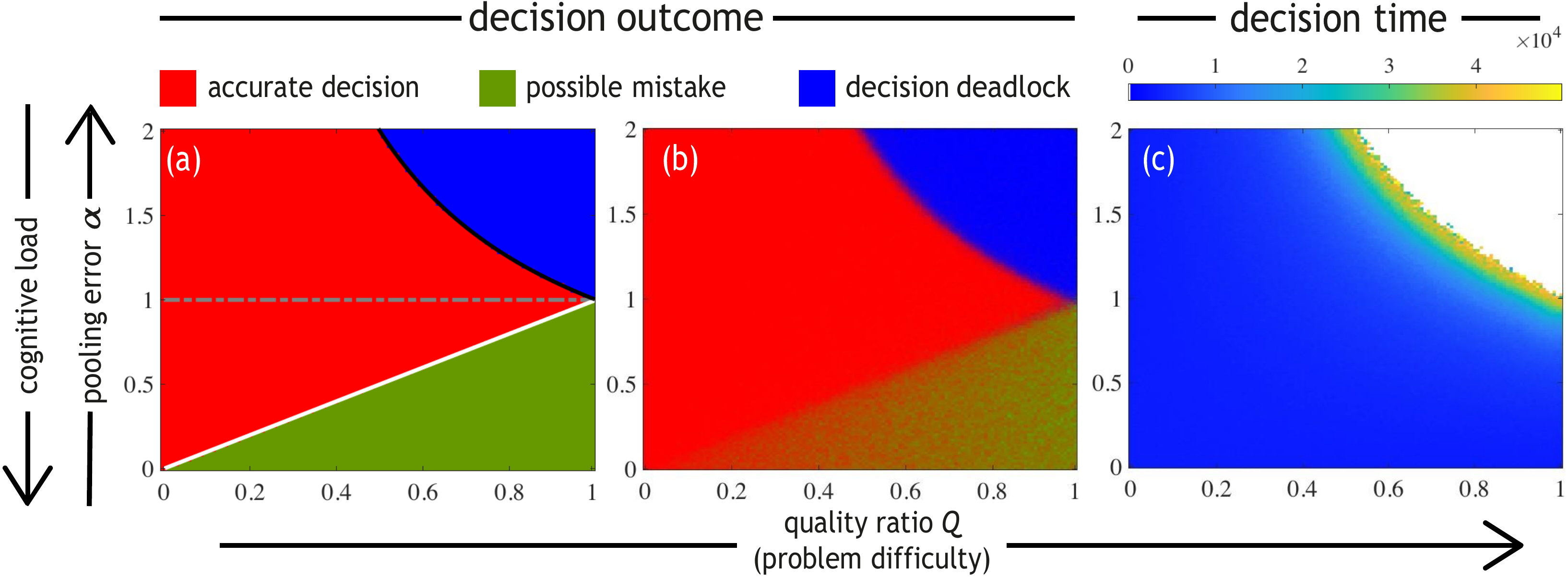}
        \caption{Stability diagrams and convergence time. In panel (a) we report the stability diagram of the mean-field model~\eqref{eq:odea2} as a function of the quality ratio $Q$ and the pooling error $\alpha$ (inversely proportional to the cognitive cost). The parameter space is divided into three regions: in the red region the population makes accurate collective decisions for any initial condition; in the blue region the population remains locked at indecision with agents' opinions fluctuating between the two options; in the green region, a consensus for either alternative is possible depending on the initial conditions, therefore mistakes are possible. The three regions are separated by the curves $\alpha=Q$ (white curve) and $\alpha=1/Q$ (black curve). Panels (b-c) show the results from simulations of $N=500$ agents interacting on a fully connected (all-to-all) network. For each couple $(Q,\alpha)$, we perform $100$ independent simulations with random initial configurations (i.e. the initial number of agents with opinion $A$ is uniformly drawn in $[0,N]$ at each run). In (b), the RGB color of each pixel is computed by assigning to the three values R, G, and B a value equal to the proportion of simulations that terminated at $n_A=500$, $n_A=0$, and $0<n_A<500$, respectively. In (c), the colormap shows the average number of timesteps needed to reach a consensus, i.e., $n_A=500$ or $n_B=500$, as a function of $Q$ and $\alpha$. The white region in the top-right corner indicates the absence of data as no simulations reached a consensus for either option.}
\label{fig:stability01complete}
\end{figure*}

Fig.\,\ref{fig:stability01complete}a shows the stability diagram of the mean-field system of Eq.\,\eqref{eq:odea2} as a function of the parameters $\alpha$ and $Q$. The parameter space is divided into three regions, determined by the stability of the system equilibria. The three regions are delimited by the curves $\alpha=Q$ (white curve) and $\alpha=1/Q$ (black curve), and correspond to the three stability cases described above in this section (see Appendix~\ref{sec:complete} for the analytical derivation of such curves). The large red region corresponds to model parameters by which the population correctly chooses the opinion with the highest quality for any initial conditions (case~(ii) above, equilibrium $\check{a}^*=0$ is unstable and $\hat{a}^*=1$ is stable). In the blue region, the population remains undecided, composed of two subpopulations, each with a different opinion (case~(iii) above, equilibria $\check{a}^*=0$ and $\hat{a}^*=1$ are unstable and $0<\tilde{a}^*<1$ is stable). In the green region, the population can converge to the best or the worst option depending on the initial condition, therefore there is the possibility the population may make a collective mistake (case~(i) above, equilibria $\check{a}^*=0$ and $\hat{a}^*=1$ are stable and $0<\tilde{a}^*<1$ is unstable). These mean-field predictions match well with the results shown in Fig.\,\ref{fig:stability01complete}b, obtained from the numerical simulation of a population of $N=500$ agents interacting on a complete graph. The outcome of the decision is color-coded in the RGB space, colouring each pixel with an RGB color where the red value $\text{R}\in[0,1]$ corresponds to the proportion of simulations terminating with accurate decisions ($n_A=N$), the green value $\text{G}\in[0,1]$ to the proportion of simulations terminating with incorrect decisions ($n_A=0$), and the blue value $\text{B}\in[0,1]$ to the proportion of simulations terminating without a consensus decision ($0<n_A<N$) after $50\,000$ timesteps.

Figs.\,\ref{fig:stability01complete}a-b show that only when the agents have a high pooling error ($\alpha > 1$), corresponding to a low cognitive load, the population can remain deadlocked at indecision. This happens in the blue region which increases in size as the cognitive load decreases (i.e., increasing pooling error $\alpha$) or the decision problem difficulty increases (i.e., increasing quality ratio $Q$). This is caused by agents that are rarely capable of changing their opinion based on the real distribution of opinions in the population. Interestingly, despite the poor skills of the agents, the population can either select the best option and make an accurate decision or remain undecided but it never selects the inferior option with the lower quality.
Differently, decision deadlocks do not occur for any cognitive level higher or equal than the voter model, i.e., $\alpha \le 1$, however, decision mistakes can be made. For any given pooling error $\alpha < 1$ (i.e., relatively high cognitive level), as the decision problem becomes harder (i.e., the quality ratio $Q$ increases), the possibility of making a mistake increases. The collective mistake occurs when the system is in the bistability (green) region and moves to a state below the unstable equilibrium, $a<\tilde{a}^*$, for example, due to random fluctuations. While the stability diagram (Fig.\,\ref{fig:stability01complete}a) only shows the region of bistability, the simulation results also show that decision mistakes are more likely to happen for harder decision problems (high $Q$).

An additional result that may not be obvious is that the red region in Figs.\,\ref{fig:stability01complete}a-b -- which corresponds to reliably accurate decisions -- is maximized for intermediate cognitive load levels (i.e., $\alpha=1$ shown as a dashed horizontal gray line) and it reduces as the agent cognitive load increases. However, we can appreciate a more complete picture of the system dynamics by also analyzing the decision time, namely the number of steps required for the population to reach a complete consensus (i.e., all agents have opinion A or all agents have opinion B). Given this definition, we can not compute the decision time when the system remains undecided and agents committed to both options coexist (blue region). Fig.~\ref{fig:stability01complete}c shows the timesteps needed to reach a consensus for a population of $N=500$ simulated agents interacting on a complete graph. (Note that the top-right corner of the colormap of Fig.~\ref{fig:stability01complete}c, which coincides with the blue region in panel b, has no decision time data because the system never reaches a consensus, therefore we left it colorless, in white.) The comparison of Figs.~\ref{fig:stability01complete}b and~\ref{fig:stability01complete}c shows the existence of a speed-accuracy trade-off, in which the speed to make a decision is traded with the collective accuracy, as it has been already documented in previous work \cite{Marshall2006,Valentini2015aamas,Valentini2016,Daniels2021}. Therefore, increasing agents' cognitive load allows the population to quickly reach a consensus at the cost of possible decision mistakes (here shown as a wider green region in Figs.~\ref{fig:stability01complete}a-b).

\section{Heterogeneous mean-field analysis}
\label{sec:HMF}

In this section, we relax the assumption of full connectivity and instead hypothesize that each agent can only exchange her opinion with a limited number of peers, i.e., with her neighbors. Thus, we can represent each agent as a node of a network connected through edges to a subset of other agents from which she can receive and send information. The network of connections is described by the adjacency matrix $\mathbf{A}$, such that $A_{ij}=1$ if and only if agents $i$ and $j$ are connected, and $A_{ij}=0$ otherwise. We also assume the network to be undirected, that is $A_{ij}=A_{ji}$, simple, that is at most one edge can connect two nodes, and connected, that is starting from any node, there is a sequence of edges that allows reaching any other node. 

When agent $i$ is selected, she pools information from her neighbors to change her opinion. Agent $i$ has $k_i=\sum_jA_{ij}$ neighbors, of which $n_{i,A}$ have opinion $A$ and $n_{i,B}$ have opinion $B$, thus, $k_i= n_{i,A}+n_{i,B}$. The number of $i$'s neighbors $k_i$ corresponds to the degree of the node $i$.
Analogous to Eq.\,\eqref{eq:defnBdies}, but assuming a sparse network, the votes expressed by the neighbors of $i$ in favor of options $A$ and $B$ are, respectively,
\begin{equation}
\begin{split}
n_{i,A}^\# = \frac{Q_A n_{i,A}/k_i}{Q_A n_{i,A}/k_i+Q_B  n_{i,B}/k_i} &\qquad \text{and} \\ 
n_{i,B}^\# = \frac{Q_B n_{i,B}/k_i}{Q_A n_{i,A}/k_i+Q_B  n_{i,B}/k_i}\, .
\end{split}
\label{eq:defni}
\end{equation}
As further described in Appendix~\ref{sec:appHMF}, once agent $i$ interacts with her neighbors, she adopts opinion $A$ or $B$ with probability $P_\alpha(n_{i,A}^\#)$ or $P_\alpha(n_{i,B}^\#)$, respectively, as we also recall that $P_\alpha(n_{i,A}^\#)=1-P_\alpha(n_{i,B}^\#)$.

By assuming the Heterogeneous Mean Field hypothesis~\cite{PSV2001,CPSV2007} to be valid, we hypothesize that all nodes with the same degree exhibit the same behavior. Therefore, we can define $A_k$ (resp. $B_k$) as the number of agents, i.e., nodes, with degree $k$ and opinion $A$ (resp. opinion $B$), and $N_k$ as the total number of agents with degree $k$. Hence, $A_k+B_k=N_k$ for all $k$. In a similar way to the mean-field analysis of Sec.~\ref{sec:mf}, we define $a_k=A_k/N_k$ (resp. $b_k=B_k/N_k$) as the proportion of agents having opinion $A$ (resp. $B$) among all agents with degree $k$; hence, for all $k$ we have $a_k+b_k=1$.

\begin{figure}[ht!]
    \centering
    \begin{tabular}{c}
        \includegraphics[width=0.5\textwidth]{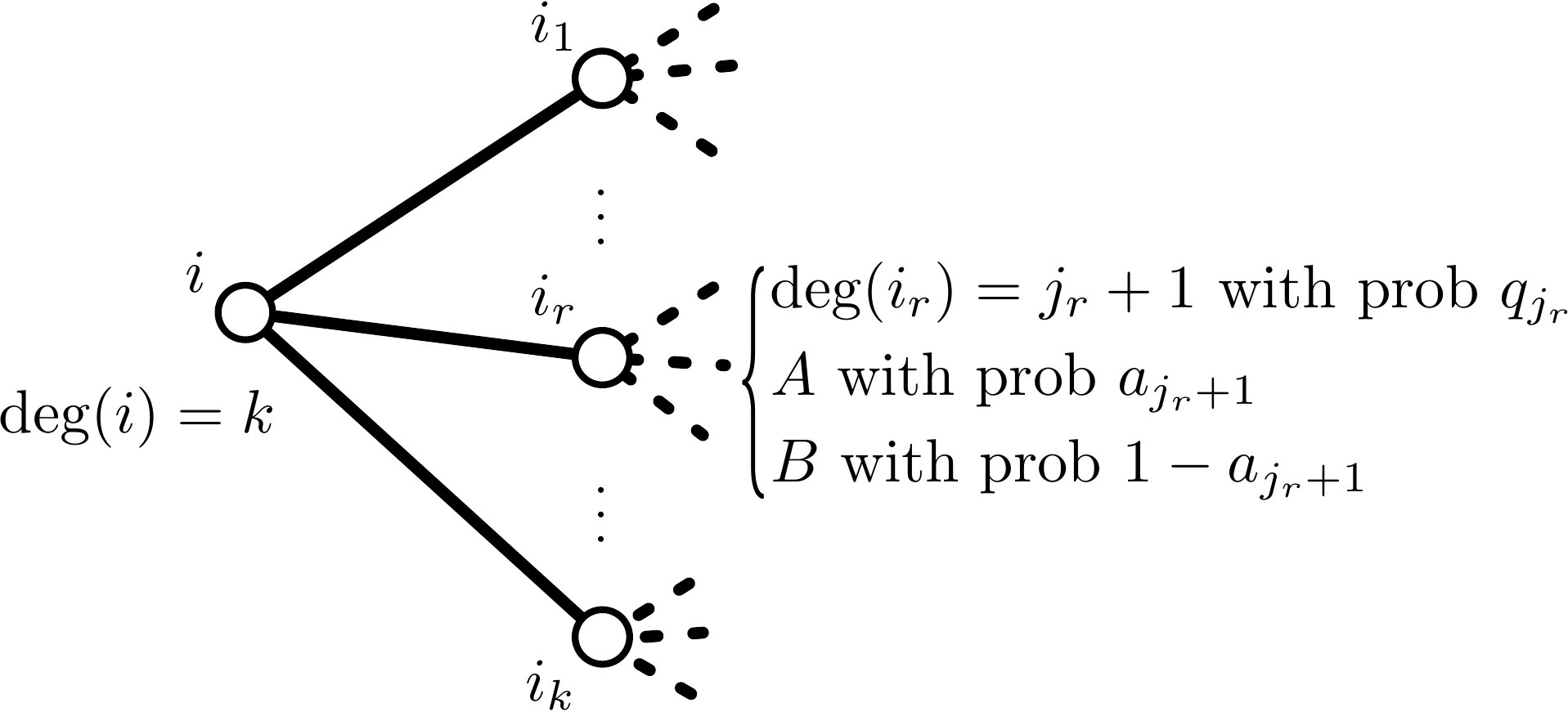} 
    \end{tabular}
        \caption{Schematic representation of the probabilities involved in the heterogeneous mean-field computations. The focal node $i$ has $k$ neighbors (degree $k$), denoted by $i_1,\dots,i_r,\dots,i_k$. Each neighbor, e.g., node $i_r$, has an excess degree $j_r$ with probability $q_{j_r}$ and therefore degree $j_r+1$. With probability $a_{j_r+1}$, she has opinion $A$, and therefore with probability $1-a_{j_r+1}$, she has opinion $B$.}
        \label{fig:HMFscheme}
\end{figure}

Let us now describe the time evolution of $a_k$ for a generic $k$. The proportion $a_k$ increases when an agent with degree $k$ and opinion $B$ changes her opinion to $A$, or it decreases when an agent with degree $k$ and opinion $A$ changes her opinion to $B$. To compute the frequency of these events, we compute $n_{i, A}^\#$ and $n_{i,B}^\#$ given in Eq.~\eqref{eq:defni} under the HMF assumption. The HMF theory originated in epidemiology, which uses the concept of {\em excess degree}~\cite{Newman} to compute the infection probability of a focal agent. The excess degree is the number of neighbors that a neighbor of the focal agent has, without considering the focal agent. In other words, the excess degree of an agent is its number of neighbors minus one (see Fig.\,\ref{fig:HMFscheme}).
We define $p_k$ as the probability that a uniformly random chosen node has degree $k$, and $q_k$ as the probability for a node to have excess degree equal to $k$, which we can compute as
\begin{equation}
\label{eq:excess}
q_k = \frac{(k+1)p_{k+1}}{\langle k \rangle} \quad \forall k\geq 0\, ,
\end{equation}
where $\langle k \rangle=\sum_k kp_k$ is the average node degree, and trivially $\sum_k q_k=1$. Let us consider a generic focal agent $i$ (or, equivalently, focal node $i$) with degree $k$ and opinion $B$ (see Fig.~\ref{fig:HMFscheme}); let $q_{j_1}$ be the probability that an agent $i_1$, connected to the focal agent $i$, has excess degree $j_1\geq 0$. Let $a_{j_1+1}$ be the probability that agent $i_1$ has opinion $A$, and $b_{j_1+1}=1-a_{j_1+1}$ the probability she has opinion $B$. By considering all the $k$ agents connected with the focal agent $i$ we can conclude that $q_{j_1}\dots q_{j_k}$ determines the joint probability that each agent reachable from any of the $k$ edges emerging from the former agent, has excess degree $j_1,\dots,j_k$. We can then define $\pi_{k,\omega}$ as the probability that $\omega$ agents among the $k$ ones have opinion $B$ and thus $k-\omega$ agents have opinion $A$. Therefore, the term $\omega$ is a linear combination of the products of $a_{j_m+1}$ and $(1-a_{j_m+1})$, with $m=1,\dots,k$. Hence, we can compute the weighted proportion of agents with opinion $A$ as $n_{i,A}^\#= (k-\omega)/[k-\omega+Q\omega]$, and this event happens with probability $q_{j_1}\dots q_{j_k}\pi_{k,\omega}$. In conclusion, agent $i$ with opinion $B$ can change opinion with probability $q_{j_1}\dots q_{j_k}\pi_{k,\omega}P_\alpha\left((k-\omega)/[k-\omega+Q\omega]\right)$. See an explanatory example for $k=2$ in Appendix~\ref{sec:appHMF}.

In a similar way, we can compute the decrease rate of agents with degree $k$ and opinion $A$. In this case, the argument of the function $P_\alpha$ is $\frac{Q_B \omega}{Q_A(k-\omega)+Q_B \omega}=\frac{Q\omega}{k-\omega+Q \omega}$, that is, the weighted proportion of agents with opinion $B$ assuming that $\omega$ agents among them have opinion $B$. Note that, because $P_\alpha(n_{A}^\#)+P_\alpha(n_{B}^\#)=1$, we have that 
\begin{equation*}
\begin{split}
P_\alpha\left(\frac{Q\omega}{k-\omega+Q \omega}\right) &=1-P_\alpha\left(1-\frac{Q\omega}{k-\omega+Q \omega}\right) \\
&=1- P_\alpha\left(\frac{k-\omega}{k-\omega+Q \omega}\right)\, .
\end{split}
\end{equation*}
By combining these equations and following a series of simplifications described in Appendix~\ref{sec:appHMF}, we describe the change in time of the proportion $a_k$ as 
 \begin{equation}
\label{eq:dakdt2}
\frac{da_k}{dt} = -a_k+\sum_{j_1,\dots,j_k}q_{j_1}\dots q_{j_k} \sum_{\omega=0}^k \pi_{k,\omega} P_\alpha\left( \frac{k-\omega}{k-\omega +\omega Q}\right)\, .
\end{equation}

To analyze the dynamics and equilibria of the system, we define
\begin{equation}
\label{eq:amean}
 \langle a\rangle := \sum_{j\geq 0} q_j a_{j+1}\, ,
\end{equation}
and by combining it with Eq.\,\eqref{eq:dakdt2}, we obtain 
\begin{equation}
\begin{split}
\label{eq:dakdt3b}
\frac{d\langle a \rangle}{dt}=-\langle a \rangle+\sum_k q_k \sum_{\omega=0}^{k} \binom{k+1}{\omega} \langle a\rangle^{k+1-\omega}\left(1-\langle a\rangle\right)^\omega \times \\
P_\alpha\left( \frac{k+1-\omega}{k+1-\omega +\omega Q}\right):=f_\alpha^{(hmf)}(\langle a\rangle)\, .
\end{split}
\end{equation}

By imposing $f_\alpha^{(hmf)}(\langle a^*\rangle)=0$, we can find (up to) three equilibria $\langle a^*\rangle$. There is the equilibrium $\langle \check{a}^*\rangle=0$, where $a_k=0$ for all $k$, which corresponds to the system with no agent having opinion $A$ (thus, all agents have opinion $B$). There is the equilibrium $\langle \hat{a}^* \rangle=1$, where $a_k=1$ for all $k$, which corresponds to the system with all agents having opinion $A$. There is also a third equilibrium $0<\langle \tilde{a}^*\rangle<1$, which has a non-trivial mathematical expression and exists only for a certain range of parameter values. This equilibrium corresponds to the coexistence of agents with opinions $A$ and $B$.

The stability of the three equilibria can be studied by analyzing the sign of the derivative of $f_\alpha^{(hmf)}$. By following the steps described in Appendix~\ref{sec:appHMF}, the derivatives evaluated at the equilibria $\langle \check{a}^*\rangle=0$ and $\langle \hat{a}^*\rangle=1$ are, respectively,
\begin{equation}
\begin{array}{ll}
\label{eq:fprime01}
(f_\alpha^{(hmf)})^\prime(0) = &-1+\sum_k q_k  (k+1) P_\alpha\left( \frac{1}{1+k Q}\right) \quad \text{and} \\ (f_\alpha^{(hmf)})^\prime(1) = &-1+\sum_k q_k (k+1)P_\alpha\left( \frac{Q}{k +Q}\right)\,.
\end{array}
\end{equation}
Therefore, we can appreciate that the stability of both equilibria is not only determined by the parameters $\alpha$ and $Q$, as in the mean-field case of Sec.~\ref{sec:mf}, but also by the network structure, via the probability of excess degree $q_k$.

\subsection{Scale-free networks}
\label{ssec:scale-free}

The heterogeneous mean-field analysis allows us to study the influence of the network topology on group dynamics. Here, we consider scale-free networks with a degree distribution that follows a power law with exponent $\gamma >2$ and minimum degree $k_{min}$.  Therefore, the probability that a uniformly random chosen node has degree $k$ is $p_k=c_\gamma/k^\gamma$ (where  $c_\gamma := \left(\sum_{k\geq k_{min}}1/k^\gamma\right)^{-1}>0$ is a normalizing constant). The excess degree probability is $q_k = \frac{1}{\langle k \rangle} \frac{c_\gamma}{(k+1)^{\gamma-1}}$, and the average degree is $\langle k \rangle=\sum_{k\geq k_{min}} k c_\gamma/k^\gamma$. In our simulations, we build scale-free networks of size $N=500$ using the configuration model~\cite{Newman}, except for $k_{min}=1$, where building connected networks of a given size through the configuration model is hard; hence to study such a case ($k_{min}=1$), as reported in the Supplementary Material, we used the Simon model~\cite{simon1955,Newman}.

\begin{figure*}[ht!]
    \centering
    \includegraphics[width=\textwidth]{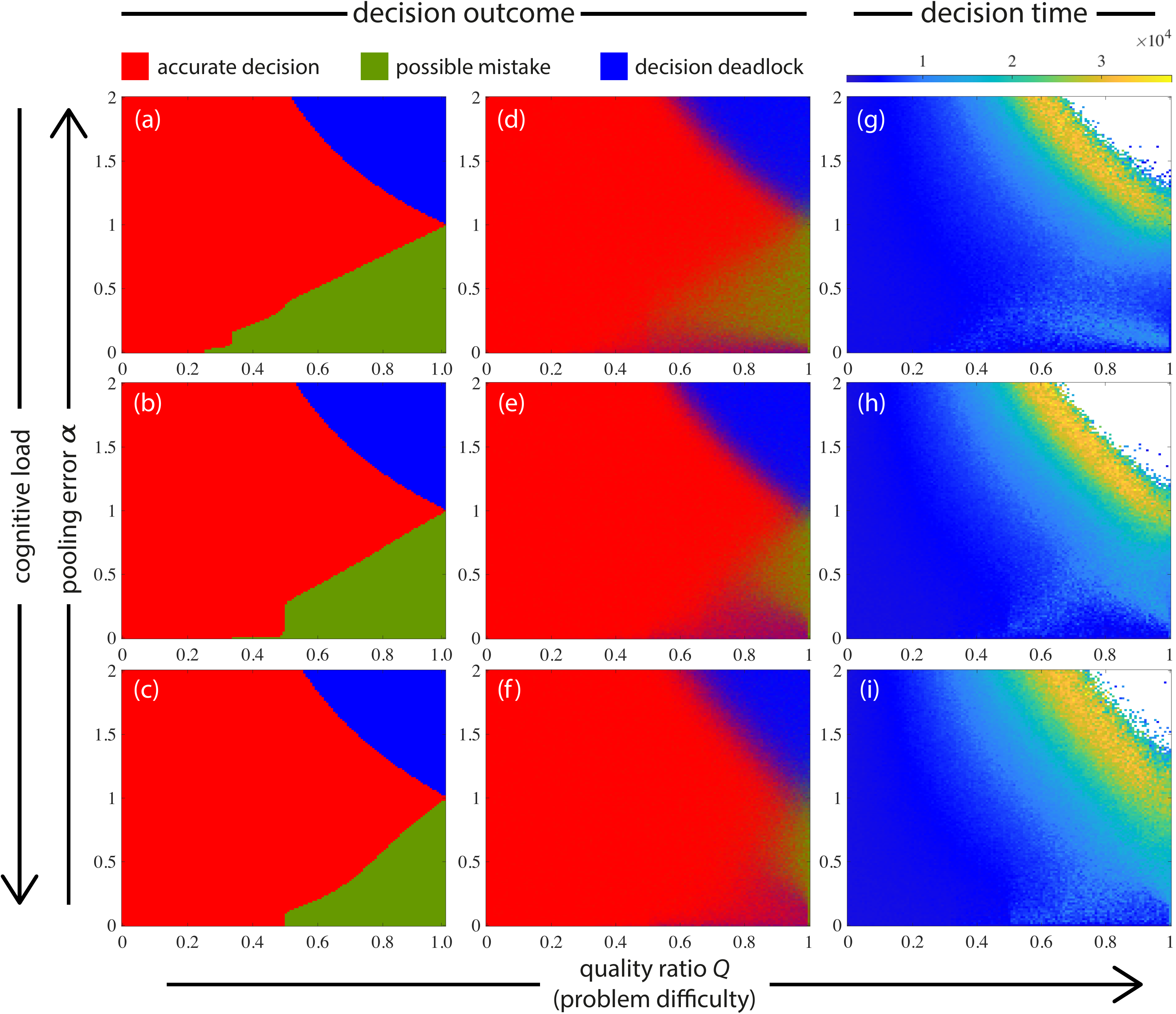}
        \caption{Stability diagrams (panels a-c), decision outcome from multiagent simulations (panels d-f) and convergence time (panels g-i) for collective decision-making on scale-free networks with $k_{min}=2$ as a function of the pooling error $\alpha$ and quality ratio $Q$. We present the results for three values of the exponent $\gamma$ regulating network connectivity: top row $\gamma=2.2$,  central row $\gamma=2.6$, bottom row $\gamma=3.1$. (a-c)~Left column panels show the convergence diagram of the mean-field model~\eqref{eq:dakdt3b}. The parameter space is divided into the same three regions of Fig.\,\ref{fig:stability01complete}a using the same color code. (d-i)~Central and right column panels show the results of simulations (100 independent runs for each $(Q,\alpha)$ configuration) of $N=500$ agents interacting on a scale-free network with random initial configurations (i.e., $n_A(t=0) \sim \mathcal{U}(0,N)$) for $50\,000$ time steps. (d-f)~Central column panels show the outcome of the collective decision-making process using the same RGB color code as Fig.\,\ref{fig:stability01complete}b. (g-i)~Right column panels show the average number of timesteps needed to reach a consensus, i.e., $n_A=500$ or $n_B=500$. The top-right white region indicates the absence of data, as the system never reaches a consensus.} 
        \label{fig:stability01scalefreeGamma}
\end{figure*}

\begin{figure*}[ht!]
    \centering
    \includegraphics[width=\textwidth]{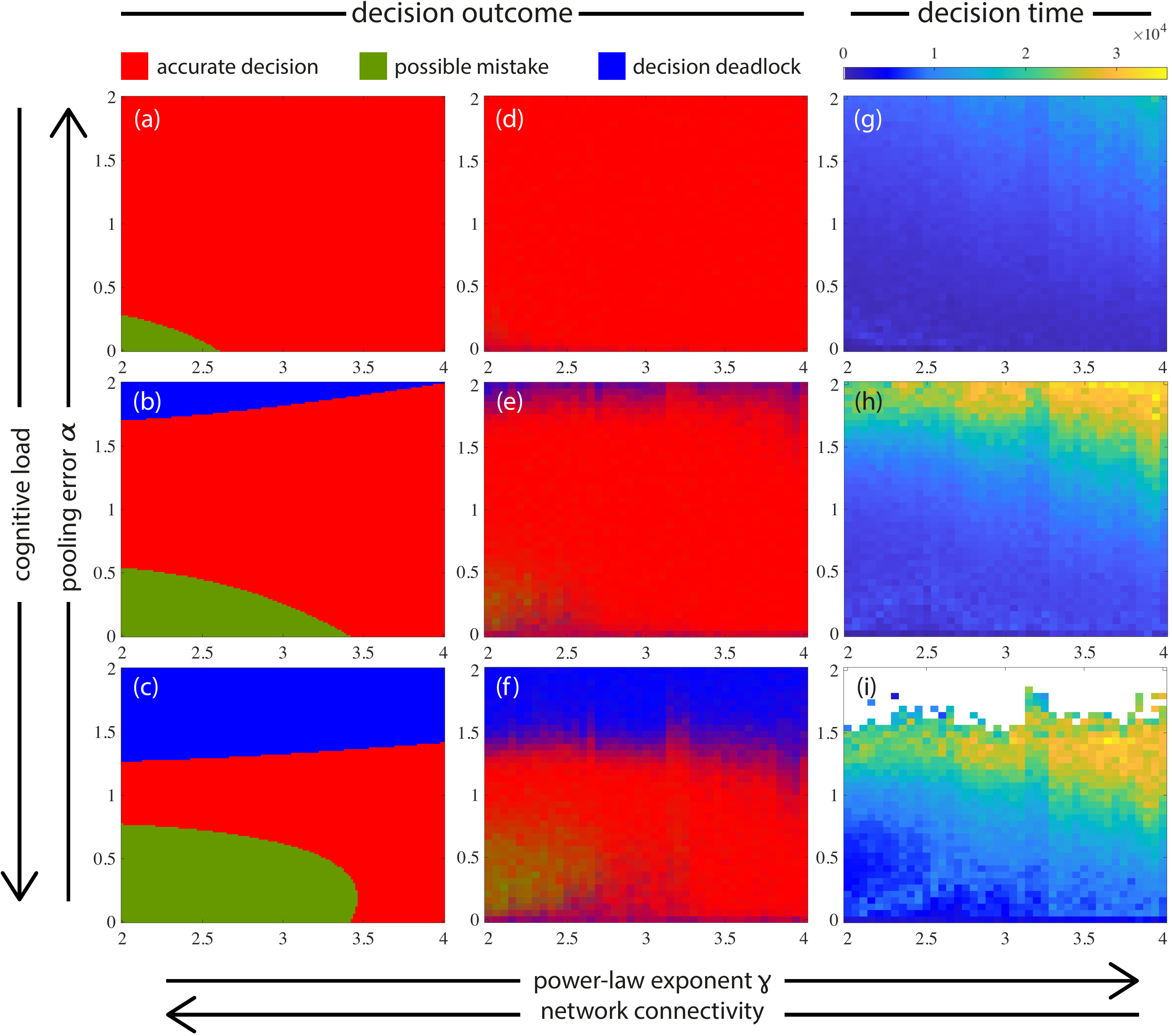}
    \caption{Stability diagrams (panels a-c), decision outcome from multiagent simulations (panels d-f) and convergence time (panels g-i) for collective decision-making on scale-free networks with $k_{min}=2$ as a function of the pooling error $\alpha$ and network's power-law exponent $\gamma$. We report results for three values of the quality ratio $Q=Q_B/Q_A$ which encodes the decision problem difficulty: top row $Q=0.5$ (easy problem), central row $Q=0.8$ (medium problem), bottom row $Q=0.9$ (difficult problem), with $Q_A=1$. Color code and experimental design are the same as the one described in the caption of Fig.\,\ref{fig:stability01scalefreeGamma}.}
        \label{fig:stability01scalefreeQ}
\end{figure*}

By using the same color code of Fig.\,\ref{fig:stability01complete}a, Figs.\,\ref{fig:stability01scalefreeGamma}(a-c) show the stability diagrams for $\gamma \in \{2.2,2.6,3.1\}$ and $k_{min}=2$ as a function of the parameters pooling error $\alpha$ and quality ratio $Q$, defining the individual cognitive load and the decision-problem difficulty, respectively. 
We can appreciate that, as the exponent $\gamma$ increases, the size of the red region increases and the size of the green region decreases (on the other hand, the change in the blue region is unnoticeable). While in the green region the collective decision depends on the initial system condition (thus opening the group to possible errors), in the red region the population cannot make mistakes but only makes accurate collective decisions for any initial condition. 
The same conclusion can be reached by observing the stability diagrams of Fig.\,\ref{fig:stability01scalefreeQ}, in which the problem difficulty is fixed (quality ratio $Q \in \{0.5,0.8,0.9\}$) and the stability regions are computed as a function of $\alpha$ and $\gamma$.
Therefore, the results of Figs.\,\ref{fig:stability01scalefreeGamma} and~\ref{fig:stability01scalefreeQ} suggest that when opinions are exchanged in networks with high $\gamma$, the population is more accurate in collective decision-making.

The effect of an improved group accuracy for higher $\gamma$ is even more evident in the numerical simulation results, shown in  Figs.\ref{fig:stability01scalefreeGamma}(d-f) and~\ref{fig:stability01scalefreeQ}(d-f), where the red region expands to an even larger parameter space than the (green) region of bistability.
This qualitative change of the group dynamics with $\gamma$ only happens for $k_{min} \le 2$ as shown in Figs.\,\ref{fig:stability01scalefreeGamma} and~\ref{fig:stability01scalefreeQ} for $k_{min} = 2$, and in Supplementary Figs.\,\ref{fig:scalefreeGamma-kmin1} and \ref{fig:scalefreeQ-kmin1} for $k_{min} = 1$. Instead, the results in Supplementary Figs.\,\ref{fig:scalefreeGamma-kmin3} and \ref{fig:scalefreeQ-kmin3} for $k_{min} = 3$ show dynamics that are (almost) independent of $\gamma$ for the mean-field (shown clearly in the $\gamma$-$\alpha$ plots by the horizontal green region of panels a-c) and have small $\gamma$-dependent improvement of accuracy in the numerical simulations (panels d-f). These results suggest that low network connectivity (here linked to a lower minimum degree $k_{min}$) can improve accuracy.

As $\gamma$ increases, the probability of having nodes with a large degree decreases as $1/k^\gamma$ and most of the nodes have a very low degree (about $k_{min}$ neighbors), also indicated by a decreasing average degree $\langle k \rangle$~\cite{Newman}. Higher $\gamma$ is also associated with a shortening of the network average shorted path $\langle \ell\rangle$, a measure of global connectivity~\cite{Newman}.
Therefore, these results suggest that collective accuracy can improve as connectivity decreases, i.e., in networks that are sparse with nodes with mostly low degrees. The counter-intuitive nature of this result -- groups perform better communicating on sparse networks -- can be explained by complementing the analysis with the decision time. In fact, panels (g)-(i) of Fig.\,\ref{fig:stability01scalefreeQ} show that as the probability of having nodes with a large degree decreases (i.e., increasing $\gamma$), the average decision time slightly increases. This result supports and extends beyond the case $\alpha=1$ the results presented in~\cite{Sood2008}. Therefore, once again, improved accuracy is coupled with slower decisions, and in this case the speed-accuracy trade-off can be regulated by both the individual cognitive load (pooling error $\alpha$) and the network connectivity (exponent $\gamma$).

\subsection{$2m$-ring networks}
\label{ssec:2mring}
\begin{figure*}[ht!]
    \centering
    \includegraphics[width=\textwidth]{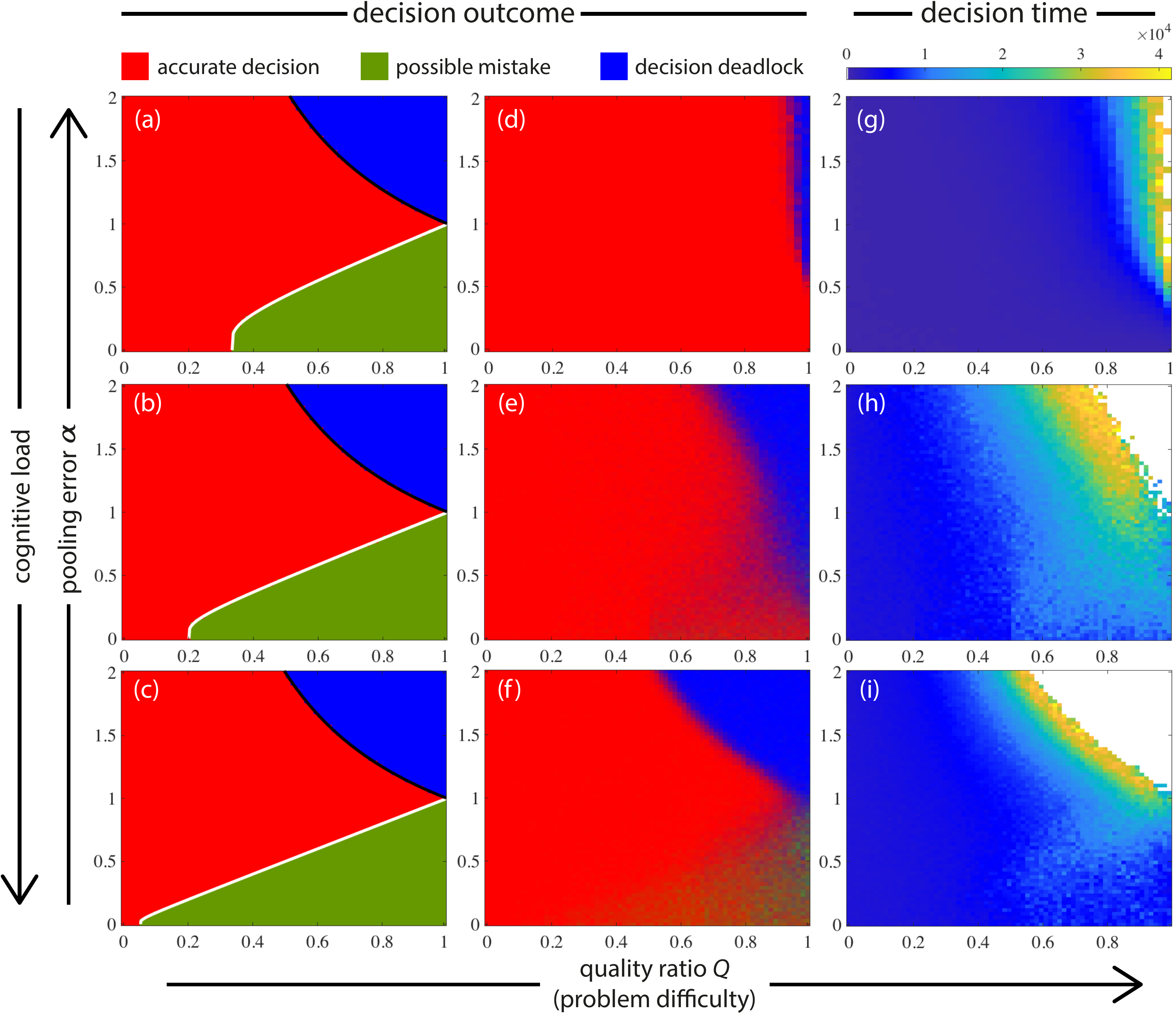} 
        \caption{Stability diagrams (panels a-c), decision outcome from multiagent simulations (panels d-f) and convergence time (panels g-i) for collective decision-making on $2m$-regular ring networks as a function of the pooling error $\alpha$ and the quality ratio $Q$. We present the results for three values of the parameter $m$ regulating network connectivity: top row $m=2$,  central row $m=3$, bottom row $m=10$. Left column panels -- i.e., (a), (d), and (g) -- show the convergence diagram of the HMF model~\eqref{eq:dakdt3b}. Color code and experimental design are the same as the one described in the caption of Fig.\,\ref{fig:stability01scalefreeGamma}.}
        \label{fig:stability01mreg}
\end{figure*}

As a cautionary note, we also want to highlight the limitations of the HMF approach. While the model predictions are confirmed by agent-based simulations on the tested scale-free networks (Figs.\,\ref{fig:stability01scalefreeGamma} and~\ref{fig:stability01scalefreeQ}), the model is not able to predict accurately the dynamics of a population interacting on $2m$-regular graphs notably in networks with small values of $m$. In particular, we investigate the dynamics of nodes interacting on ring networks where all nodes are connected to their first $m$ neighbor nodes on the ``left'' and $m$ on the ``right'', in such a way all nodes have the same degree $k=2m$. In Appendix~\ref{sec:complete}, we study the heterogeneous mean-field model for the case of $2m$-regular ring networks and derive the equations describing the bifurcation points that determine the stability changes of the system.
Figs.\,\ref{fig:stability01mreg}(a-c) show stability diagrams for $m \in \{2,3,10\}$ that are qualitatively similar to the ones computed for the other types of networks (complete graph and scale-free networks). However, Figs.\,\ref{fig:stability01mreg}(d-f) show that the results of the numerical simulations only partially agree with the theory predictions. The agreement improves as the connectivity increases ($m=10$, panel f), however, the dynamics for low values of $m$ are different. In particular, we can appreciate that the agent-based simulations can make accurate decisions (red region) for a much larger range of parameters $Q$ and $\alpha$ than what the theory predicted. 

\begin{figure}[ht!]
    \centering
    \includegraphics[width=0.4\textwidth]{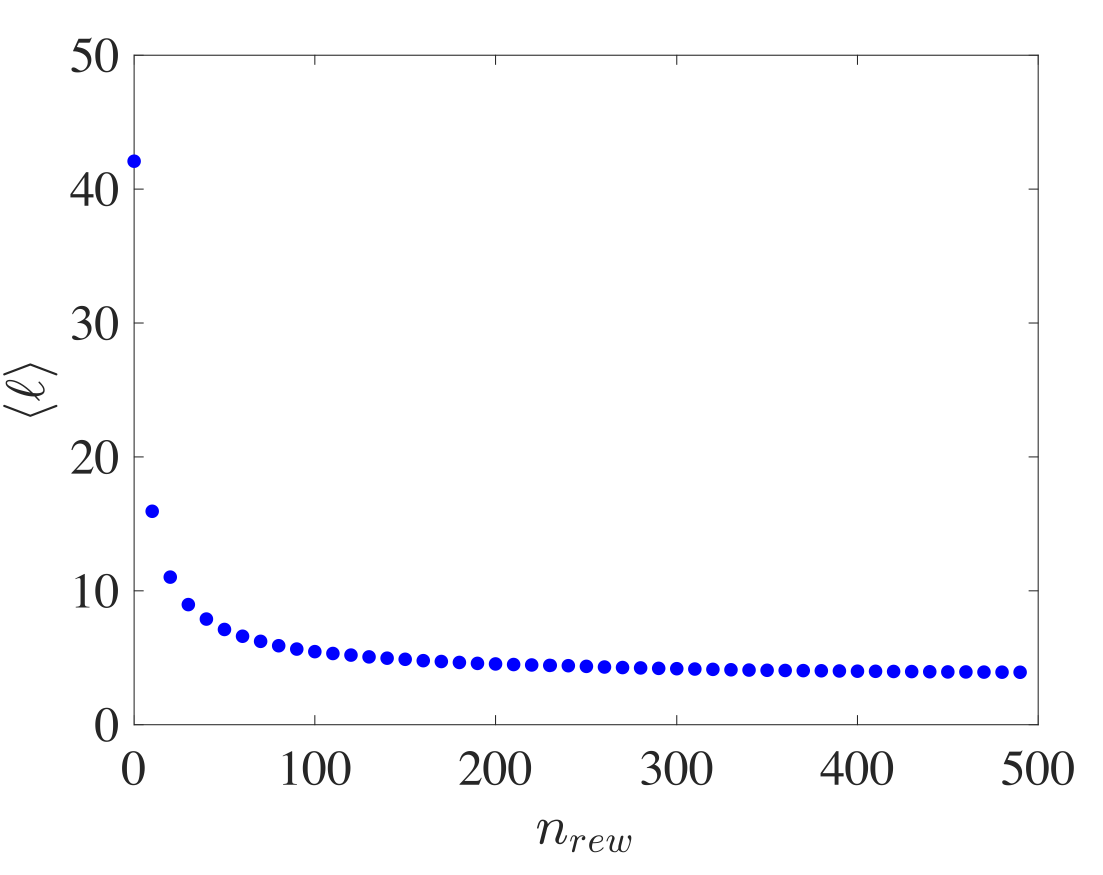}
    \caption{Average shortest path $\langle \ell\rangle$ (y-axis) as a function of the number of criss-cross rewiring $n_{new}$ (x-axis), starting from a regular $6$-ring network of $N=500$ nodes.}
        \label{fig:avel}
\end{figure}

One possible cause of this discrepancy between the mean-field model dynamics and the simulation results can be the model assumption of a well-mixed system, which is not satisfied in $2m$-ring networks. Our intuition is further supported by the analysis of the option dynamics on Erd\H{o}s-R\'enyi graphs, reported in Appendix~\ref{sec:ERgraph}, which show a good agreement between the HMF model's stability diagrams and the multiagent simulations. Therefore, we applied a degree preserving rewiring process to the $2m$-ring networks (see Appendix~\ref{sec:complete}) to reduce the average shortest path and study its impact on the population dynamics. More precisely, we consider a $6$-regular ring, i.e. $m=3$, made of $N=500$ nodes, and rewire a subset of the network edges.
In this way, all nodes keep $2m$ neighbors and, as shown in Fig.\,\ref{fig:avel}, only the average shortest path $\langle \ell\rangle :=\frac{\sum_{i\neq j} \ell_{ij}}{N(N-1)}$ reduces as the rewiring increases (where $\ell_{ij}$ is the shortest path among nodes $i$ and $j$).

\begin{figure*}[ht!]
    \centering
    \includegraphics[width=\textwidth]{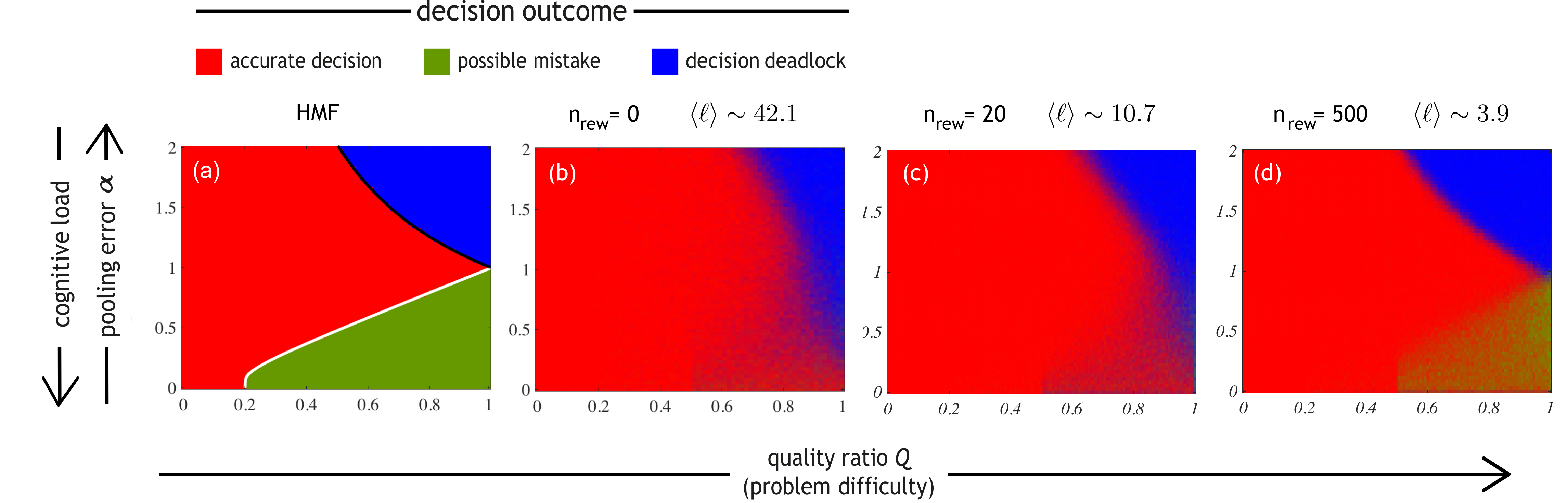}
    \caption{Comparison of the HMF model stability diagram (panel a) with the results of the numerical simulations (panels b-d). We study the impact of the criss-cross rewiring in $2m$-ring networks, with $m=3$ (i.e., $6$-ring). (a)~Stability diagram of the HMF model (this is the same as Fig.\,\ref{fig:stability01mreg}d). (b-d)~Decision outcome of the numerical simulations using the same color code and experimental design as the ones described in the caption of Fig.\,\ref{fig:stability01scalefreeGamma}d. On top of each panel, we indicate how many edge-rewiring steps ($n_{new}$) we compute and the corresponding average shortest path $\langle \ell\rangle$, i.e., in panel~(b) no rewiring was done and corresponds to Fig.\,\ref{fig:stability01mreg}e), in panel~(c) we do $n_{rew}=20$ edge-rewiring steps, and in panel~(d) we did $n_{rew}=500$ edge-rewiring steps.}
        \label{fig:ringrew}
\end{figure*}

By comparing the results of the HMF model and of the numerical multiagent simulations on a $6$-ring network with increasing rewiring, we can appreciate a good match between model and simulations as the number of rewiring increases (Fig.\,\ref{fig:ringrew}). Note that all rewired networks have the same degree distribution $p_k=2m=6$ and thus the HMF model's equilibria are the same. With the average shortest path shortening, the similarity between the model and the numerical simulation results increases.

Despite the current HFM model cannot completely grasp the dynamics on $2m$-regular ring networks, the simulation results on this type of networks (Figs.\,\ref{fig:stability01mreg}d-f) confirm our intuition that collective decision-making accuracy increases as the network connectivity reduces. Population operating on networks with low $m$ have consistently a higher accuracy. Interestingly, for $2m$-regular ring networks, it seems that there is no trade-off between decision speed and accuracy but, instead, rings with low $m$ enable both quick and accurate decisions.

\section{Conclusions}
\label{sec:conclusions}

Modeling collective decision-making processes can provide useful insights into the understanding of the living world and design new decision protocols for more efficient group decisions. The voter model and variations of them, thanks to their mathematical simplicity, have been effectively used to model decision-making at every level of biological complexity, from ecological dynamics of plant communities \cite{Zillio2005} to coordinated motion in fish \cite{Jhawar2020} to house-hunting in honeybees \cite{Reina:PRE:2017} to human group dynamics \cite{Castellano2009, Redner2019}. In the best-of-$n$ problem, the group has to select the best option among a discrete set of $n$ alternatives and these models describe how opinions spread from one individual to another through voting interactions on a social network.
This study presents a model capable of generalizing a set of popular existing voter-like models for collective decision-making in the best-of-2 scenario. Our analysis focuses on understanding the impact of two control parameters -- the pooling error and the network connectivity -- on the collective performance in terms of decision accuracy and time.

Both pooling error and the network connectivity regulate a speed-accuracy trade-off. By reducing the pooling error, and therefore demanding a higher cognitive effort from the individuals to correctly process the social information, the population can make quicker and more democratic decisions, however with reduced accuracy due to the more frequent selection of the option with the lowest quality, compared with models with higher pooling error and thus lower individual cognitive effort. Instead, by reducing network connectivity, and therefore reducing the average number of neighbors of each individual, collective accuracy is improved at the cost of higher decision time. 
These results improve our understanding of the role of individual costs and network connectivity in collective decision-making.

By measuring the losses, and even the benefits, of reducing computation and communication costs, our analysis can be useful to support the design of autonomous robot systems capable of operating without human supervision. Reduced costs can save energy, money, and, in general, resources both at design and run time, i.e., designing robots with simpler circuitry can be cheaper and consuming less energy in social interactions can improve efficiency. Erroneous computations and limited connectivity can not only be cheaper but also increase performance in terms of decision accuracy (due to a longer deliberation time during which the decision, rather than being rushed, is more accurately made). Recent previous results also observed that there are certain conditions where reduced connectivity between the agents of the group -- robots, animals, or humans -- can give important group-level advantages, e.g., better responsiveness to environmental changes \cite{Talamali:SciRobot:2021, Aust:ANTS:2022, Mateo2019}, evading a predator or avoiding dangers \cite{Sosna2019, Rahmani2020}, or generate higher cultural diversity and innovations \cite{Derex2016,Gershenson2015}. We believe that the results of our analysis can also have important implications in the study and design of group decision-making in human societies, which can be biased and manipulated through targeted interventions on how information is exchanged and aggregated in the social networks \cite{Momennejad2022,Bak-Coleman2021}.

Despite abstracting components of the process, the type of analysis proposed in our study can give useful predictions. Future work can look into extending the analysis to measure the impact of different aspects -- e.g., average degree, shortest average path -- characterizing the different types of network topologies -- e.g., Erd\H{o}s-R\'{e}nyi random graphs, Barab\'asi-Albert's scale-free networks, Watts-Strogatz's small-word network, and random geometric graphs. We also believe that an interesting extension of the work could apply the present analysis to study the dynamics of heterogeneous populations, comprising individuals that follow different voting rules or that have different levels of conformism with others (e.g., populations comprising stubborn agents \cite{mobilia2003,Reina:CommPhy:2023,Njougouo:WIVACE:2023}).

\section*{Acknowledgement}
We would like to thank Luca Gallo for the useful discussions we had during the preparation of the final manuscript.

\bibliography{biblio}

\newpage
\appendix 
\onecolumngrid

\begin{center}
\large{\textbf{Supplementary Material.\\Studying speed-accuracy trade-offs in best-of-n collective decision-making\\through heterogeneous mean-field modelling}}
\end{center}

\section{The mean-field model}
\label{sec:compeltegraph}
In this appendix, we first indicate the steps needed to deduce the mean-field system of Eq.\,\eqref{eq:odea2} (in the main text) and then we find the equilibria of the system and study their stability.

Let us introduce the proportion of agents with opinion $A$ (resp. $B$), $a(t)=n_A(t)/N$ (resp. $b(t)=n_B(t)/N$), hence $a(t)+b(t)=1$. The proportion of agents with opinion $A$ increases because agents with opinion $B$ change their minds and adopt opinion $A$, or decreases if agents with opinion $A$ adopt opinion $B$; therefore, we can write the change of $a$ in a small time interval $dt \rightarrow 0$ as
\begin{equation}
\label{eq:odea}
\frac{da}{dt} = bP_\alpha\left(a^\#\right) - aP_\alpha\left(b^\#\right) \, .
\end{equation}
We recall the quantities $n_A^\#$ and $n_B^\#$ defined in Eq.\,\eqref{eq:defnBdies} in the main text, representing the votes expressed for option $A$ and $B$, respectively, which are weighted by the quality. Therefore, we can define the weighted proportions
\begin{equation}
\label{eq:defnadies}
a^\# = \frac{a}{a+Q b} \quad \text{ and } \quad b^\#= \frac{Q b}{a+Q b}\, ,
\end{equation}
where $Q$ is the ratio $Q_B/Q_A$. Given the functional form of $P_\alpha$ given in Eq.\,\eqref{eq:Px}, we can conclude that
\begin{equation*}
P_\alpha\left(a^\#\right) + P_\alpha\left(b^\#\right)=1\, .
\end{equation*}
Hence, we can derive the mean-field model of Eq.\,\eqref{eq:odea2} as
\begin{equation*}
\frac{da}{dt} = (1-a)P_\alpha\left(a^\#\right) - a\left(1-P_\alpha\left(a^\#\right)\right) =P_\alpha\left( \frac{a}{a(1-Q)+Q}\right) - a=:f_\alpha(a)\, .
\end{equation*}

The equilibria of this system are determined by the zeros of $f_\alpha(a)$, namely we are looking for values $a^*\in[0,1]$ such that
\begin{equation}
\label{eq:odeazero}
f_\alpha(a^*)=0\,, \quad \text{i.e.,} \quad P_\alpha\left( \frac{a^*}{a^*(1-Q)+Q}\right) - a^*=0\, .
\end{equation}
The equilibrium stability is determined by the sign of the derivative computed on the equilibrium, $f'(a^*)$. To simplify the analysis, let us introduce a new variable
\begin{equation}
\label{eq:xa}
 x=\frac{a}{a(1-Q)+Q}\, .
\end{equation}
Observe that $x$ is well defined, indeed $a(1-Q)+Q\neq 0$ for $a\in [0,1]$ and moreover $x=0$ if $a=0$ and $x=1$ if $a=1$. In conclusion, Eq.\,\eqref{eq:xa} defines a bijective map from $[0,1]$ into $[0,1]$. By inverting the relation~\eqref{eq:xa} we can write
\begin{equation}
\label{eq:ax}
 a=\frac{xQ}{1-x(1-Q)}\, ,
\end{equation}
hence solving Eq.~\eqref{eq:odeazero} is equivalent to solve
\begin{equation}
\label{eq:odeazero2}
P_\alpha\left( x^*\right) = \frac{x^*Q}{1-x^*(1-Q)}\, ,
\end{equation}
that is to determine the intersections between the function $P_\alpha\left( x\right)$ and the hyperbola $g(x)=\frac{xQ}{1-x(1-Q)}$. Two solutions are trivially found: $\check{x}^*=0$ and $\hat{x}^*=1$. However, for some choice of the parameters $Q$ and $\alpha$, a third solution $\tilde{x}^*$ can also exist.

To find the intersections between the two functions, let us define $\Delta(x)=P(x)-g(x)$, and search for which values $\Delta(x)=0$. As indicated above, for $x=0$ and $x=1$, we have $\Delta(0)=\Delta(1)=0$. To determine the existence of (at least) a third root let us consider the derivative of $\Delta(x)$ at $x=0$ and $x=1$. A straightforward computation returns $\Delta^\prime(0)=\alpha -Q$ and $\Delta^\prime(1)=\alpha-1/Q$. A sufficient condition to have a third solution is thus $\Delta^\prime(0)>0$ and $\Delta^\prime(1)>0$ or $\Delta^\prime(0)<0$ and $\Delta^\prime(1)<0$. The reason is that the function $\Delta(x)$ is continuous and if it approaches both $x=0$ and $x=1$ with increasing (or both decreasing) derivatives, it should cross (at least once) the $0$ line at some point.

Let us now consider the stability of the three equilibria. The equilibrium $\check{x}^*=0$ is stable if and only if $\Delta^\prime(0)<0$, namely if $\alpha <Q$. On the other hand, the equilibrium $\hat{x}^*=1$ is stable if and only if $\Delta^\prime(1)<0$, namely $\alpha<1/Q$. From Eq.\,\eqref{eq:ax} one can obtain the value of the variable $a$ given $x$, we can thus draw the following conclusions:
\begin{itemize}
 \item Let $Q>1$:
 \begin{itemize}
\item if $\alpha < 1/Q$, then we also have $\alpha < Q$, thus $\check{a}^*=0$ and $\hat{a}^*=1$ are {\em stable} equilibria, and the third equilibrium $0<\tilde{a}^*<1$ exists but it is {\em unstable};
\item if $1/Q<\alpha < Q$, then $\check{a}^*=0$ is {\em stable}, $\hat{a}^*=1$ is {\em unstable}, and the third equilibrium $0<\tilde{a}^*<1$ does not exist;
\item if $Q<\alpha$, then $\check{a}^*=0$ and $\hat{a}^*=1$ are {\em unstable} equilibria, and the third equilibrium $0<\tilde{a}^*<1$ exists and is {\em stable}.
\end{itemize}
 \item Let $Q<1$:
  \begin{itemize}
\item if $\alpha < Q$, then we also have $\alpha <1/Q$, thus $\check{a}^*=0$ and $\hat{a}^*=1$ are {\em stable} equilibria, and the third equilibrium $0<\tilde{a}^*<1$ exists but is {\em unstable};
\item if $Q<\alpha < 1/Q$, then $\check{a}^*=0$ is {\em unstable}, $\hat{a}^*=1$ is {\em stable}, and the third equilibrium $0<\tilde{a}^*<1$ does not exist;
\item if $Q<\alpha$, then $\check{a}^*=0$ and $\hat{a}^*=1$ are {\em unstable} equilibria, and the third equilibrium $0<\tilde{a}^*<1$ exists and is {\em stable}.
\end{itemize}
\end{itemize}

\section{The heterogeneous mean-field model}
\label{sec:appHMF}

In this section, we present the detailed computation needed to derive through heterogeneous mean-field theory Eqs.\,\eqref{eq:dakdt2} and~\eqref{eq:dakdt3b}, presented in the main text. Let us assume that agents are connected via a network and they can exchange opinions only with neighbors to which they are directly corrected. Given an agent $i$, her neighbors are defined as the nodes $j$ for which $A_{ij}=1$, where $\mathbf{A}$ is the $N\times N$ adjacency matrix. Observe that $A_{ij}=0$ if agents $i$ and $j$ are not connected and therefore cannot directly exchange opinions.

Let us introduce the quantities $n_{i,A}$ and $n_{i,B}$ that indicate the number of neighbors of agent $i$ with opinion $A$ and $B$, respectively. Formally, we can define them as 
\begin{equation}
 n_{i,A}=\sum_j A_{ij}\hat{A}_j \quad \text{ and } \quad n_{i,B}=\sum_j A_{ij}\hat{B}_j \, ,
\end{equation}
where $\hat{A}_j=1$ (resp. $\hat{B}_j=1$) if agent $j$ has opinion $A$ (resp. $B$), and zero otherwise.

An agent $i$ with opinion $A$ (resp. $B$) changes her opinion to $B$ (resp. $A$) with probability defined by the nonlinear function $P_\alpha(n_{i,B}^\#)$ (resp. $P_\alpha(n_{i,A}^\#)$) presented in Eq.\,\eqref{eq:Px}, with argument the weighted proportion $n_{i,B}^\#$ (resp. $n_{i,A}^\#$) of $i$'s neighbors with opinion $B$ (resp. $A$). The weights of $n_{i,A}^\#$ and $n_{i,B}^\#$ are proportional to the quality of opinions $A$ and $B$, respectively, and can be mathematically defined as
 \begin{equation}
n_{i,A}^\# = \frac{Q_A n_{i,A}}{Q_A n_{i,A}+Q_B  n_{i,B}} \quad \text{and} \quad n_{i,B}^\# = \frac{Q_B n_{i,B}}{Q_A n_{i,A}+Q_B  n_{i,B}}\,.
\label{eq:weightedProps}
\end{equation}
We recall that $k_i=\sum_jA_{ij}$ is the degree of the node $i$ and trivially $k_i= n_{i,A}+n_{i,B}$. Let us observe that by defining $Q=Q_B/Q_A$ we can rewrite the previous relations as:
\begin{equation}
 n_{i,A}^\# = \frac{n_{i,A}}{n_{i,A}+Q  n_{i,B}}=\frac{n_{i,A}}{(1-Q)n_{i,A}+Q k_i}=\frac{n_{i,A}/k_i}{(1-Q)n_{i,A}/k_i+Q} \quad \text{and} \quad  n_{i,B}^\# = 1- n_{i,A}^\# \, ,
\end{equation}
namely Eq.\,\eqref{eq:defni} in the main text.

Let us now assume the validity of the Heterogeneous Mean Field hypothesis (HMF)~\cite{PSV2001,CPSV2007} and let thus aggregate agents according to their opinion and degree, namely we define $A_k$ and $B_k$ to be the number of agents with degree $k$ and opinion $A$ or $B$, respectively. Then, setting $N_k$ to be the number of agents with degree $k$, we have $A_k+B_k=N_k$ for all $k$. Let us introduce $a_k=A_k/N_k$ and $b_k=B_k/N_k$ as the proportion of agents with degree $k$ and opinion $A$ or $B$, respectively. The goal is to express the probability of changing opinion by using the HMF.

Let us consider an agent $i$ with opinion $B$ and assume she has $k$ neighbors; and we want to compute the probability that $\omega$ neighbors have opinion $B$ and $k-\omega$ opinion $A$ (with $\omega\in\{0,\dots,k\}$), so that we can compute the weighted proportions of Eq.\,\eqref{eq:weightedProps} as $n_{i,A}^\#= (k-\omega)/[k-\omega+Q\omega]$ and $n_{i,B}^\#= \omega Q/[k-\omega+Q\omega]$. In the spirit of the HMF hypothesis, we determine the probability that a node has degree $k^\prime$ by only knowing that it is connected to a node with degree $k$; the latter is given by the {\em excess degree} $q_{k^\prime}$, namely $q_{k^\prime} ={({k^\prime}+1)p_{{k^\prime}+1}}/{\langle k \rangle}$, where $p_k$ is the proportion of nodes with degree $k$ and $\langle k \rangle$ the average network degree.

\subsubsection*{Example for $k=2$}

Before computing the formula for a general degree $k$, let us present an example for $k=2$ which helps us to explain our reasoning. Assume the focal agent $i$ has degree $k_i=2$ and the two neighbors have excess degree $j_1\geq 0$ and $j_2\geq 0$, then there are three possible cases:
\begin{itemize}
 \item Both neighbors have opinion $A$. This happens with probability
 \begin{equation*}
 q_{j_1}a_{j_1+1}q_{j_2}a_{j_2+1}\, ,
\end{equation*}
\item Both neighbors have opinion $B$. This happens with probability
\begin{equation*}
 q_{j_1}(1-a_{j_1+1})q_{j_2}(1-a_{j_2+1})\, .
\end{equation*}
\item One neighbor has opinion $A$ and one neighbor has opinion $B$. This happens with probability
\begin{equation*}
 q_{j_1}a_{j_1+1}q_{j_2}(1-a_{j_2+1})+ q_{j_1}(1-a_{j_1+1})q_{j_2}a_{j_2}\, .
\end{equation*}
\end{itemize}
Let us define $\pi_{2,\omega}$ be the probability that $\omega\in\{0,1,2\}$ agents have opinion $B$ and thus $2-\omega$ opinion $A$. Then the previous three cases can be summarized into a single formula
\begin{equation*}
 q_{j_1}q_{j_2}\pi_{2,\omega}\,, \quad\forall \omega\in\{0,1,2\}\, .
\end{equation*}

We can now compute the probability that the focal agent $i$ changes her opinion to $A$:
\begin{itemize}
 \item In the case both neighbors have opinion $A$,
\begin{equation*}
 q_{j_1}q_{j_2}\pi_{2,0} P_\alpha\left(\frac{2-0}{2-0+0Q}\right)= q_{j_1}q_{j_2}\pi_{2,0} P_\alpha\left(1\right)\, .
\end{equation*}
\item In the case both neighbors have opinion $B$,
\begin{equation*}
 q_{j_1}q_{j_2}\pi_{2,2} P_\alpha\left(\frac{2-2}{2-2+2Q}\right)= q_{j_1}q_{j_2}\pi_{2,2} P_\alpha\left(0\right)\, .
\end{equation*}
\item In the case one of the two neighbors has opinion $A$ the other has opinion $B$, 
\begin{equation*}
 q_{j_1}q_{j_2}\pi_{2,1} P_\alpha\left(\frac{2-1}{2-1+1Q}\right)= q_{j_1}q_{j_2}\pi_{2,1} P_\alpha\left(\frac{1}{1+Q}\right)\, .
\end{equation*}
\end{itemize}

\subsubsection*{The general case}

The reasoning presented in the example for $k=2$ can be repeated for a general $k$. Then $q_{j_1}\dots q_{j_k}$ evaluates the joint probability that each node reachable from any of the $k$ edges emerging from the focal node, has excess degree $j_1,\dots,j_k$. We can define $\pi_{k,\omega}$, to be the probability that $\omega$ nodes among the $k$ neighbors have opinion $B$ and thus $k-\omega$ opinion $A$. Therefore, the term $\pi_{k,\omega}$ is a linear combination of products of $a_{j_m+1}$ and $(1-a_{j_m+1})$, with $m=1,\dots,k$. Finally, the probability $q_{j_1}\dots q_{j_k}\pi_{k,\omega}$ is multiplied by the function $P_\alpha$ with argument the weighted proportion of agents with opinion $A$ or $B$, that is $\frac{k-\omega}{k-\omega+Q \omega}$ or $\frac{\omega Q}{k-\omega+Q \omega}$.

Let us observe that because of property~\eqref{eq:PaPB} (i.e., $P(n_A^\#)+P(n_B^\#)=1$), we have that
\begin{equation*}
P_\alpha\left(\frac{Q\omega}{k-\omega+Q \omega}\right)=1-P_\alpha\left(1-\frac{Q\omega}{k-\omega+Q \omega}\right)=1- P_\alpha\left(\frac{k-\omega}{k-\omega+Q \omega}\right)\, .
\end{equation*}
Thus, we can wrap together the above expressions and obtain the time evolution of $a_k$ for a generic degree $k$:
\begin{equation}
\label{eq:dakdt}
\frac{da_k}{dt} = (1-a_k)\sum_{j_1,\dots,j_k}q_{j_1}\dots q_{j_k} \sum_{\omega=0}^k \pi_{k,\omega}P_\alpha\left( \frac{k-\omega}{k-\omega +\omega Q}\right)-a_k\sum_{j_1,\dots,j_k}q_{j_1}\dots q_{j_k} \sum_{\omega=0}^k \pi_{k,\omega}\left[1- P_\alpha\left( \frac{k-\omega}{k-\omega +\omega Q}\right)\right]\, .
\end{equation}
Let us explain each term on the right-hand side. The leftmost term, $(1-a_k)$, is the probability that the focal agent has degree $k$ and does not have opinion $A$, she hence has opinion $B$. The term $q_{j_1}\dots q_{j_k}$ evaluates the joint probability that each node reachable from any of the $k$ edges emerging from the focal node, has excess degree $j_1,\dots,j_k$; the sum $\sum_{j_1,\dots,j_k}$ allows to consider all the possibilities. For a given choice of $j_1,\dots,j_k$, the next term, $\pi_{k,\omega}$, determines the probability that $\omega$ nodes among the $k$ ones have opinion $B$ and thus $k-\omega$ have opinion $A$. The sum $\sum_{\omega=0}^k$ allows to consider all the possibilities from $\omega=0$, all agents have opinion $A$, to $\omega=k$, all agents have opinion $B$. Finally, the term $P_\alpha\left( \frac{k-\omega}{k-\omega +\omega Q}\right)$ is the probability the focal agent with opinion $B$ changes her mind because there are $k-\omega$ agents with opinion $A$ and $\omega$ agents with opinion $B$. The remaining terms denote the opposite process where the selected agent has opinion $A$, with probability $a_k$, and she changes opinion after an interaction with her neighbors with opinion $B$. As already observed, we used the property~\eqref{eq:PaPB} for the function $P_\alpha$ to rewrite the rightmost term.

Eq.~\eqref{eq:dakdt} can be split into four parts
\begin{eqnarray*}
\frac{da_k}{dt} &=& \sum_{j_1,\dots,j_k}q_{j_1}\dots q_{j_k} \sum_{\omega=0}^k \pi_{k,\omega}P_\alpha\left( \frac{k-\omega}{k-\omega +\omega Q}\right)-a_k\sum_{j_1,\dots,j_k}q_{j_1}\dots q_{j_k} \sum_{\omega=0}^k \pi_{k,\omega}P_\alpha\left( \frac{k-\omega}{k-\omega +\omega Q}\right)+\\&-&a_k\sum_{j_1,\dots,j_k}q_{j_1}\dots q_{j_k} \sum_{\omega=0}^k \pi_{k,\omega}+a_k\sum_{j_1,\dots,j_k}q_{j_1}\dots q_{j_k} \sum_{\omega=0}^k \pi_{k,\omega}P_\alpha\left( \frac{k-\omega}{k-\omega +\omega Q}\right)\, ,
\end{eqnarray*}
and we can observe that the rightmost terms on the first and second line do simplify each other by returning
\begin{equation}
\frac{da_k}{dt} = \sum_{j_1,\dots,j_k}q_{j_1}\dots q_{j_k} \sum_{\omega=0}^k \pi_{k,\omega}P_\alpha\left( \frac{k-\omega}{k-\omega +\omega Q}\right)-a_k\sum_{j_1,\dots,j_k}q_{j_1}\dots q_{j_k} \sum_{\omega=0}^k \pi_{k,\omega}\, .
\label{eq:hmf:dak}
\end{equation}

We trivially have $\sum_{\omega=0}^k \pi_{k,\omega}=1$ and, by assuming absence of correlations among nodes degrees, we also have $\sum_{j_1,\dots,j_k}q_{j_1}\dots q_{j_k}=\sum_{j_1} q_{j_1}\dots \sum_{j_k}q_{j_k}=1$, hence we can simplify Eq.\,\eqref{eq:hmf:dak} into:
 \begin{equation}
\label{eq:dakdt2app}
\frac{da_k}{dt} = \sum_{j_1,\dots,j_k}q_{j_1}\dots q_{j_k} \sum_{\omega=0}^k \pi_{k,\omega} P_\alpha\left( \frac{k-\omega}{k-\omega +\omega Q}\right)-a_k\, .
\end{equation}
As already indicated in the main text, we define $\langle a\rangle := \sum_j q_j a_{j+1}$. By using combinatorics and assuming probability independence, we can show that 
\begin{equation*}
 \sum_{j_1,\dots,j_k}q_{j_1}\dots q_{j_k}  \pi_{k,\omega} =  \binom{k}{\omega} \langle a\rangle^{k-\omega}\left(1-\langle a\rangle\right)^\omega\,;
\end{equation*}
the rough idea is that in $\pi_{k,\omega}$ there are $k-\omega$ events with probability $a_{j_m+1}$, thus $\omega$ with $(1-a_{j_m+1})$, and the binomial coefficient computes all possible permutations. We can thus rewrite Eq.\,\eqref{eq:dakdt2app} as
\begin{equation}
\label{eq:dakdt3}
\frac{da_k}{dt} = -a_k+ \sum_{\omega=0}^{k-1} \binom{k}{\omega} \langle a\rangle^{k-\omega}\left(1-\langle a\rangle\right)^\omega P_\alpha\left( \frac{k-\omega}{k-\omega +\omega Q}\right)\, ,
\end{equation}
where we removed from the sum the term $\omega=k$ because it contains $P_\alpha\left(0\right)=0$. By rewriting the previous equation with $k\rightarrow k+1$, and by multiplying both sides by $q_{k}$ and summing over $k$ to bring out $\langle a\rangle$ we get Eq.~\eqref{eq:dakdt3b} in the main text, namely:
\begin{eqnarray*}
\frac{d\langle a \rangle}{dt}&=&\sum_k q_k \frac{da_{k+1}}{dt} = -\sum_k q_k a_{k+1}+\sum_k q_k \sum_{\omega=0}^{k} \binom{k+1}{\omega} \langle a\rangle^{k+1-\omega}\left(1-\langle a\rangle\right)^\omega P_\alpha\left( \frac{k+1-\omega}{k+1-\omega +\omega Q}\right)\notag \\
&=&-\langle a \rangle+\sum_k q_k \sum_{\omega=0}^{k} \binom{k+1}{\omega} \langle a\rangle^{k+1-\omega}\left(1-\langle a\rangle\right)^\omega P_\alpha\left( \frac{k+1-\omega}{k+1-\omega +\omega Q}\right)\, ,
\end{eqnarray*}
where the right hand side defines the function $f_\alpha^{(hmf)}(\langle a\rangle)$.

\subsubsection*{Stability analysis}

Let us now consider the zeros of $f_\alpha^{(hmf)}(\langle a\rangle)$, hence the equilibria of the system. Because the sum over $\omega$ ranges from $\omega=0$ and $\omega=k$, and because the involved terms are of the form $\langle a\rangle^{k+1-\omega}$, they all vanish once $\langle a\rangle=0$, hence $f_\alpha^{(hmf)}(\langle a\rangle)=0$. The same holds true for $\langle a\rangle=1$, indeed
\begin{equation*}
f_\alpha^{(hmf)}(1) =  -1+\sum_k q_k\sum_{\omega=0}^{k} \binom{k+1}{\omega} \left(1-\langle a^*\rangle\right)\rvert_{\langle a^*\rangle=1}^\omega P_\alpha\left( \frac{k+1-\omega}{k+1-\omega +\omega Q}\right)=  -1+\sum_k q_{k}\, ,
\end{equation*}
where we used the fact that all the terms $\left(1-\langle a^*\rangle\right)\rvert_{\langle a^*\rangle=1}^\omega$ vanish except the one with $\omega=0$, for which we also have $\binom{k+1}{0}=1$ and $P_\alpha\left( \frac{k+1}{k+1}\right)=1$. The conclusion follows by recalling that $\sum_k q_{k}=1$.

The stability of the above equilibria can be determined by considering the derivative of $f_\alpha^{(hmf)}$ at $0$ and $1$ that is given by
\begin{equation}
\label{eq:fprime}
\begin{split}
\left(f_\alpha^{(hmf)}\right)^\prime(\langle a^*\rangle) = -1+\sum_k q_k\sum_{\omega=0}^{k} \binom{k+1}{\omega} P_\alpha\left( \frac{k+1-\omega}{k+1-\omega +\omega Q}\right) \times\\\left[ (k+1-\omega)\langle a^*\rangle^{k-\omega}\left(1-\langle a^*\rangle\right)^\omega - \omega\langle a^*\rangle^{k+1-\omega}\left(1-\langle a^*\rangle\right)^{\omega-1}\right] \notag\, ,
\end{split}
\end{equation}
hence
\begin{equation}
\label{eq:df0}
\left(f_\alpha^{(hmf)}\right)^\prime(0) = -1+\sum_k q_k  \binom{k+1}{k} P_\alpha\left( \frac{1}{1+k Q}\right)=-1+\sum_k q_k  (k+1) P_\alpha\left( \frac{1}{1+k Q}\right)\, ,
\end{equation}
and
\begin{eqnarray}
\label{eq:df1}
\left(f_\alpha^{(hmf)}\right)^\prime(1)&=& -1+\sum_k q_k \left[ \binom{k+1}{0} P_\alpha\left( \frac{k+1}{k+1}\right) (k+1)-\binom{k+1}{1} P_\alpha\left( \frac{k}{k +Q}\right)\right]\notag \\
&=&-1+\sum_k q_k (k+1) \left[1- P_\alpha\left( \frac{k}{k +Q}\right)\right]=-1+\sum_k q_k (k+1)P_\alpha\left( \frac{Q}{k +Q}\right) \, .
\end{eqnarray}

\begin{figure}[htp!]
    \centering
    \vspace{-2cm}
    \begin{tabular}{c}
        \includegraphics[width=0.42\textwidth]{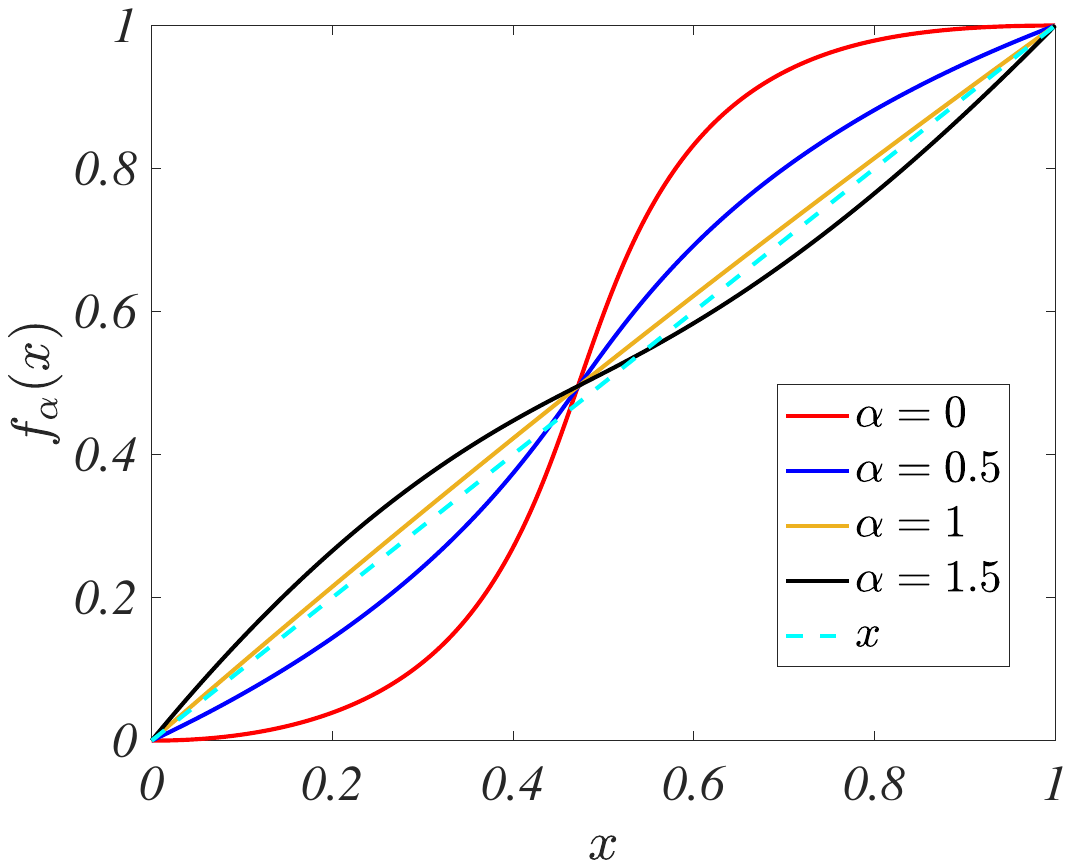} \vspace{-2.5cm}
    \end{tabular}
        \caption{The function $f_\alpha(x)=x+f_\alpha^{(hmf)}(x)$ for some values of $\alpha$ for a scale free network made by $N=400$ nodes, $\gamma = 2.2$, $k_{min}= 3$, $k_{max}=189$, the quality ratio is $Q=0.9$. The dashed line represents the identity curve and thus its intersections with the function $f_\alpha(x)$ determine the equilibria of the system.}
        \label{fig:figfHMF}
\end{figure}

In Fig.~\ref{fig:figfHMF} we report four examples of the function $f_\alpha(x)=x+f_\alpha^{(hmf)}(x)$ for four values of $\alpha$ for a scale-free network with exponent $\gamma=2.2$. Observe that, differently from Fig.\,\ref{fig:Px}, the function is smooth even for $\alpha=0$ (red curve). 
Additionally, the presence of three intersections of $f_\alpha(x)$ with the line $y=x$, hence three zeros for $f_\alpha^{(hmf)}(x)$, indicates the presence of three system equilibria. For $\alpha=1$ (yellow line), there are only two line intersections, at $x=0$ and $x=1$, indicating the existence of only two equilibria.

\subsubsection*{Limitations of the heterogeneous mean-field model}

We conclude this appendix by studying the limitation of the heterogeneous mean-field model, more precisely we look for (family of) networks for which there is a disagreement between the dynamics predicted by the HMF model and the numerical agent-based simulations. More precisely, we look at the case in which the HMF model's equilibrium $\langle \check{a}^*\rangle=0$ is unstable, i.e., Eq.~\eqref{eq:df0} is positive, and $\langle \hat{a}^*\rangle=1$ is stable, i.e., Eq.~\eqref{eq:df1} is negative. Therefore, the prediction is consistent convergence for any initial state to the latter stable equilibrium $\langle \hat{a}^*\rangle=1$; however, numerical simulations do not always terminate with all agents with opinion $A$.

For any given $Q<1$ there exists $\bar{k}$ such that $kQ>1$ for all $k\geq \bar{k}$ and thus $kQ<1$ for all $k< \bar{k}$. Hence 
\begin{equation*}
 kQ>1\Rightarrow \frac{1}{1+kQ}<\frac{1}{2}\Rightarrow P_\alpha\left(\frac{1}{1+kQ}\right) = \frac{1}{2}-\frac{1}{2}\left(1-\frac{2}{1+kQ}\right)^\alpha\, ,
\end{equation*}
and
\begin{equation*}
 kQ<1\Rightarrow \frac{1}{1+kQ}>\frac{1}{2}\Rightarrow P_\alpha\left(\frac{1}{1+kQ}\right) = \frac{1}{2}+\frac{1}{2}\left(\frac{2}{1+kQ}-1\right)^\alpha\, .
\end{equation*}
Eq.\,\eqref{eq:df0} rewrites thus
\begin{eqnarray*}
\left(f_\alpha^{(hmf)}\right)^\prime(0) &=&-1+\frac{1}{2}\sum_{k\geq \bar{k}} q_k  (k+1) \left[ 1-\left(1-\frac{2}{1+kQ}\right)^\alpha\right]+\frac{1}{2}\sum_{k< \bar{k}} q_k  (k+1) \left[ 1+\left(\frac{2}{1+kQ}-1\right)^\alpha\right]=\\ 
&=&-1+\frac{1}{2}\sum_{k} q_k  (k+1)+\frac{1}{2}\sum_{k< \bar{k}} q_k  (k+1) \left(\frac{2}{1+kQ}-1\right)^\alpha-\frac{1}{2}\sum_{k\geq  \bar{k}} q_k  (k+1) \left(1-\frac{2}{1+kQ}\right)^\alpha \, .
\end{eqnarray*}
By using the definition of $q_k$, we can compute
\begin{equation*}
 \sum_{k\geq 0} q_k  (k+1)=\sum_{k\geq 0} \frac{(k+1)p_{k+1}}{\langle k\rangle}  (k+1)=\frac{1}{\langle k\rangle}\sum_{k\geq 1} k^2 p_{k}=\frac{1}{\langle k\rangle}\sum_{k\geq 0} k^2 p_{k}=\frac{\langle k^2\rangle}{\langle k\rangle}\, ,
 \end{equation*}
 hence
 \begin{eqnarray*}
\left(f_\alpha^{(hmf)}\right)^\prime(0) =-1+\frac{1}{2}\frac{\langle k^2\rangle}{\langle k\rangle}+\frac{1}{2}\sum_{k< \bar{k}} q_k  (k+1) \left(\frac{2}{1+kQ}-1\right)^\alpha-\frac{1}{2}\sum_{k\geq  \bar{k}} q_k  (k+1) \left(1-\frac{2}{1+kQ}\right)^\alpha \, .
\end{eqnarray*}
To compute $\left(f_\alpha^{(hmf)}\right)^\prime(0)$ we observe that if $k\geq 1>Q$ then $Q/(k+Q)<1/2$ and thus we can conclude
 \begin{eqnarray*}
\left(f_\alpha^{(hmf)}\right)^\prime(1) &=&-1+q_0+\frac{1}{2}\sum_{k\geq 1} q_k  (k+1) \left[ 1-\left(1-\frac{2Q}{k+Q}\right)^\alpha\right]=\\ 
&=&-1+\frac{q_0}{2}+\frac{1}{2}\frac{\langle k^2\rangle}{\langle k\rangle}-\frac{1}{2}\sum_{k\geq 1} q_k  (k+1) \left(1-\frac{2Q}{k+Q}\right)^\alpha \, .
\end{eqnarray*}

For sake of definitiveness let us assume $1/2<Q<1$ and thus $\bar{k}=2$.  Hence $\left(f_\alpha^{(hmf)}\right)^\prime(0)$ simplifies into
\begin{eqnarray*}
\left(f_\alpha^{(hmf)}\right)^\prime(0)=-1+\frac{1}{2}\frac{\langle k^2\rangle}{\langle k\rangle}+ q_1 \left(\frac{2}{1+Q}-1\right)^\alpha-\frac{1}{2}\sum_{k\geq  2} q_k  (k+1) \left(1-\frac{2}{1+kQ}\right)^\alpha \, .
\end{eqnarray*}
Finally let us consider a $2$-ring network where, i.e., each node is connected with its two neighbors, hence $\langle k\rangle =2$, $\langle k^2\rangle =4$, $p_2=1$ and $p_k=0$ for all $k\neq 2$ and thus $q_1=1$ and $q_k=0$ if $k\neq 1$. The previous equation simplifies to give
 \begin{eqnarray*}
\left(f_\alpha^{(hmf)}\right)^\prime(0)=-1+\frac{1}{2}\frac{4}{2}+ \left(\frac{2}{1+Q}-1\right)^\alpha= \left(\frac{2}{1+Q}-1\right)^\alpha>0\, ,
\end{eqnarray*}
namely $\langle \check{a}^*\rangle=0$ is unstable under the assumption of HMF. Similarly the equation for $\left(f_\alpha^{(hmf)}\right)^\prime(1)$ rewrites
 \begin{equation*}
\left(f_\alpha^{(hmf)}\right)^\prime(1) = -1+\frac{1}{2}\frac{4}{2}-\frac{1}{2} q_1 2 \left(1-\frac{2Q}{1+Q}\right)^\alpha =- \left(\frac{1-Q}{1+Q}\right)^\alpha <0\, ,
\end{equation*}
namely $\langle \hat{a}^*\rangle=1$ is stable according to the HMF theory.

\begin{figure}[t!]
    \centering
    \begin{tabular}{c}
        \includegraphics[width=1\textwidth]{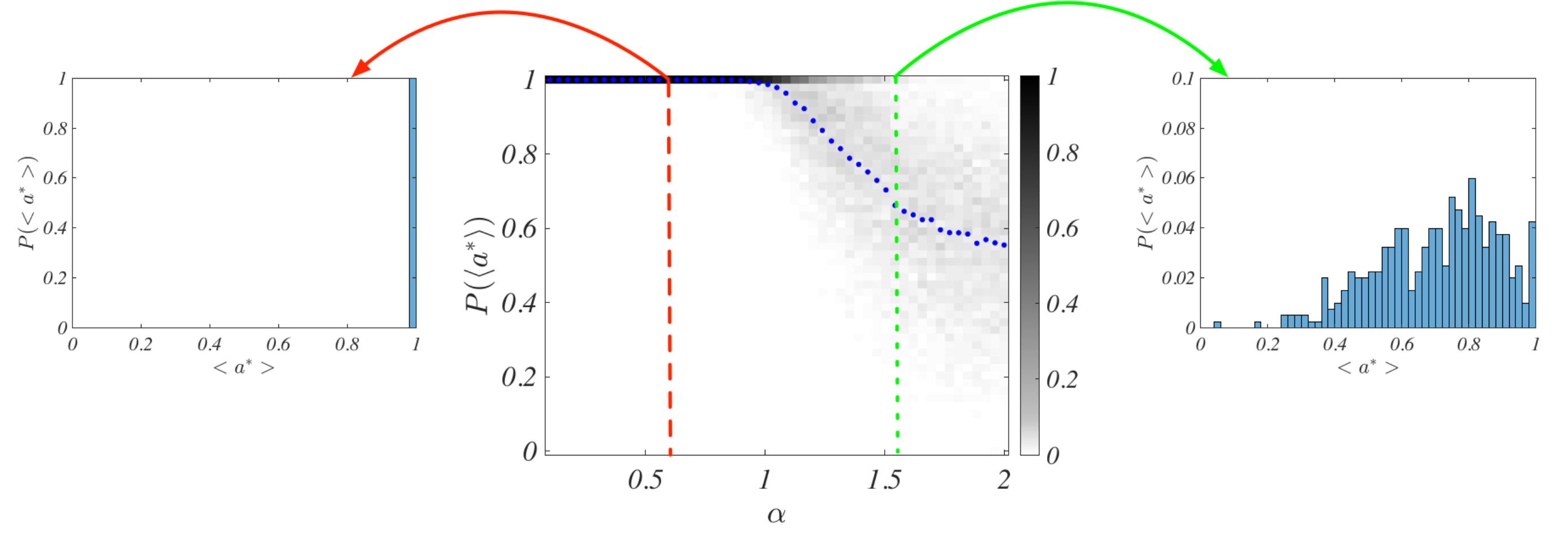}
    \end{tabular}
        \caption{Asymptotic state for the agent based model on a $2$-ring. In the main panel we report the distribution of the asymptotic state $\langle n_A^*\rangle/N$ for for $N=500$ agents interacting in a $2$-ring as a function of the parameter $\alpha$ for a fixed $Q=0.9$. For each given $\alpha$ we repeat the simulation $n_{iter}=400$ times and we then we display the probability distribution of the obtained values (the darker the higher the probability). On the left panel we report the case small $\alpha$, here $\alpha= 0.59$, and we can appreciate the fact the all the simulations returned $\langle n_A^*\rangle/N=1$; on the right panel we show the case of large $\alpha$, here $\alpha=1.5$, and we can observe a large spreading of  values about the mean (blue dots in the main panel) but also the presence of a small peak corresponding to $\langle n_A^*\rangle/N=1$ and an even smaller for $\langle n_A^*\rangle/N=0$.}
        \label{fig:ringHMF}
\end{figure}

In Fig.~\ref{fig:ringHMF} we show the results of numerical simulation of $500$ agents exchanging opinions on a $2$-ring, i.e., each agent has two neighbors. Each point is asymptotic value after $500\, 000$ time steps, of $\langle n_A^*\rangle/N$ averaged over $401$ independent simulations as a function of $\alpha$ for $Q=0.9$. Simulations have different initial conditions (initial opinions distributed differently on the network) but always with $250$ agents with opinion $A$ and $250$ agents with opinion $B$, i.e., half of the population committed to each option. One can observe that for $\alpha < 1$ the simulations converge to $\langle n_A^*\rangle/N = 1$ and thus the claim of the predictions of the HMF model are confirmed. Whereas the good match between theory and simulations is no longer valid for $\alpha >1$. Theory predicts a single stable equilibria for full consensus for $A$, while the simulated system remains locked at indecision at $0<\tilde{a}*<1$ with only a part of the agents with opinion $A$ and the rest with opinion $B$.

The different behavior for $\alpha >1$ is due to finite-size effects and can be explained as follows. Let us assume to have a ring with $N$ agents and assume that all agents but one have opinion $A$, we are interested in the possibility that also this last $B$-agent changes her mind and becomes $A$-agent, in this way the system reaches the equilibrium $\langle n_A^*\rangle /N= 1$, i.e., all $A$. This process should be compared with the one where one agent $A$ becomes a $B$-agent, the ratio of the probabilities of those two events determines the stability (or not) of the state all $A$.

The probability that the event $B\rightarrow A$ happens is the combination of the probability of selecting the $B$-agent, hence probability $1/N$, and the probability that she will receive a message from a neighbour committed to $A$. Because there is only one $B$-agent, her two neighbors have opinion $A$ and thus $P_\alpha(1)=1$. In summary, $P(B\rightarrow A)=1/N$. 

The probability that the event $A\rightarrow B$ happens is the combination of the probability of selecting one of the two agents $A$ sitting in the ring next to the unique agent $B$, hence probability $2/N$, and the probability that the selected agent $A$ will select the message from the $B$-agent. The weighted proportion of agents $B$ that are neighbors of the selected agent $A$ is $P_\alpha\left(\frac{Q}{1+Q}\right)$. Being $Q\in [0,1]$, the quantity $\frac{Q}{1+Q}$ is smaller than $1/2$ and thus $P_\alpha\left(\frac{Q}{1+Q}\right)=\frac{1}{2}-\frac{1}{2}\left(\frac{1-Q}{1+Q}\right)^\alpha$. In summary, $P(A\rightarrow B)=\frac{2}{N} \left[\frac{1}{2}-\frac{1}{2}\left(\frac{1-Q}{1+Q}\right)^\alpha\right]$.

Finally, we can compute the ratio of the probabilities for the two events
\begin{equation*}
 \frac{P(A\rightarrow B)}{P(B\rightarrow A)} = 1-\left(\frac{1-Q}{1+Q}\right)^\alpha\,.
\end{equation*}
Because $\frac{1-Q}{1+Q}<1$, we can conclude that if $\alpha<1$, then $ \frac{P(A\rightarrow B)}{P(B\rightarrow A)}$ is small and thus the system has a large probability to evolve toward a consensus for $A$, the event $B\rightarrow A$ is much more probable than $A\rightarrow B$. On the other hand, if $\alpha >1$, then the above probability approaches $1$ and thus both events are (almost) equally probable. Therefore, despite the equilibrium of a consensus for $A$ being stable, it is difficult to reach it because the two events ($A\rightarrow B$ and $B\rightarrow A$) are equally likely to happen and the system can fluctuate indefinitely.


\section{The $2m$-regular graph}
\label{sec:complete}

Let us consider now $2m$-regular graphs, $m\geq 1$, namely networks where all the nodes have the same degree $2m$. Observe that $1$-dimensional rings where each node is connected to $m$ left and $m$ right neighbors fall in this class, that however contains more general structures. By construction we trivially have $p_k=1$ if $k=2m$ and $p_k=0$ otherwise, then $\langle k\rangle = 2m$, which implies that $q_k=1$ if $k=2m-1$ and $0$ otherwise, indeed
\begin{equation*}
q_k=\frac{k+1}{\langle k \rangle} p_{k+1}=
\begin{cases}
\frac{(2m-1)+1}{2m}p_{2m} =1& \text{if $k+1=2m$}\\
0 & \text{if $k+1\neq 2m$}\, .
\end{cases} 
\end{equation*}

Therefore, the weighted average proportion of agents with opinion $A$ simply becomes $\langle a\rangle = a_{2m}$. From the definition~\eqref{eq:dakdt3b}, we can simplify the function $f_\alpha^{(hmf)}(x)$ and obtain
\begin{equation}
\label{eq:fsimp}
f_\alpha^{(hmf)}(x)= -x+\sum_{\omega=0}^{2m-1} \binom{2m}{\omega} x^{2m-\omega}\left(1-x\right)^\omega P_\alpha\left( \frac{2m-\omega}{2m-\omega+\omega Q}\right)\,.
\end{equation}
The derivatives of~\eqref{eq:fsimp} evaluated at $\check{x}^*=0$ and $\hat{x}^*=1$ are given by (see also~\eqref{eq:df0} and~\eqref{eq:df1}):
\begin{equation*}
(f_\alpha^{(hmf)})^\prime(0) = -1+2m P_\alpha\left( \frac{1}{1+(2m-1) Q}\right) \quad \text{and} \quad (f_\alpha^{(hmf)})^\prime(1) =-1+2m P_\alpha\left( \frac{Q}{2m-1+Q}\right) \, .
\end{equation*}
In conclusion, the equilibrium $\check{a}^*_{2m}=0$, i.e., all agents have opinion $B$, is stable if and only if 
\begin{equation*}
(f_\alpha^{(hmf)})^\prime(0) = -1+2m P_\alpha\left( \frac{1}{1+(2m-1) Q}\right)<0 \, ,
\end{equation*}
and similarly $\hat{a}^*_{2m}=1$, i.e., all agents have opinion $A$, is stable if and only if 
\begin{equation*}
(f_\alpha^{(hmf)})^\prime(1) = -1+2m P_\alpha\left( \frac{Q}{2m-1+Q}\right) <0 \, .
\end{equation*}

When $Q<1/(2m-1)$, then $1/[(2m-1)Q+1]>1/2$, therefore, using the definition of $P_\alpha$ we get:
\begin{equation*}
(f_\alpha^{(hmf)})^\prime(0) = -1+2m \left[\frac{1}{2}+\frac{1}{2}\left( \frac{2}{1+(2m-1) Q}-1\right)^\alpha\right] =-1+m \left[1+\left( \frac{2}{1+(2m-1) Q}-1\right)^\alpha\right] \geq -1+m\, ,
\end{equation*}
and because $m \geq 2$, the latter expression is positive for all $\alpha\geq 0$. In conclusion, the equilibrium $\check{a}^*_{2m}=0$ is unstable for all $\alpha \geq 0$ and $Q<1/(2m-1)$.

Let us now consider the case  $Q>1/(2m-1)$. By definition of $P_\alpha$ we get
\begin{equation*}
(f_\alpha^{(hmf)})^\prime(0) = -1+2m \left[\frac{1}{2}-\frac{1}{2}\left( 1-\frac{2}{1+(2m-1) Q}\right)^\alpha\right]=-1+m \left[1-\left( 1-\frac{2}{1+(2m-1) Q}\right)^\alpha\right] \, ,
\end{equation*}
and now the quantity on the right-hand side can have both signs. Let us define $\hat{\alpha}(Q)$ as the value of $\alpha$ for which the right-hand side vanishes for a fixed $Q>1/(2m-1)$, then we can straightforwardly obtain
\begin{equation}
\label{eq:hatalpha}
\hat{\alpha}(Q) = \frac{\log\left(1-\frac{1}{m}\right)}{\log\left(1-\frac{2}{1+(2m-1) Q}\right)}\, .
\end{equation}
By looking at its definition we can conclude that $\hat{\alpha}(1)=1$ and that $\hat{\alpha}(Q)\rightarrow 0$ for $Q\rightarrow 1/(2m-1)$ (from values larger than $1/(2m-1)$). Given $Q>1/(2m-1)$, then $(f_\alpha^{(hmf)})^\prime(0)>0$ for all $\alpha > \hat{\alpha}(Q)$; this means that the equilibrium $\check{a}^*_{2m}=0$ is unstable. The function $\hat{\alpha}(Q)$ is drawn in white in Figs.~\ref{fig:stability01mreg}a-c and it delimits the red region where $(f_\alpha^{(hmf)})^\prime(0)>0$ ($\check{a}^*_{2m}=0$ is unstable) and the green region where $(f_\alpha^{(hmf)})^\prime(0)<0$ ($\check{a}^*_{2m}=0$ is stable).

Let us now consider the stability of the equilibrium $\hat{a}^*_{2m}=1$. Because $m\geq 1$ and $Q<1$, we always have $Q/(2m-1+Q)<1/2$, hence by definition of $P_\alpha$ we obtain
\begin{equation}
(f_\alpha^{(hmf)})^\prime(1) = -1+2m \left[\frac{1}{2}-\frac{1}{2}\left(1- \frac{2Q}{2m-1 +Q}\right)^\alpha\right] = -1+m \left[1-\left(1- \frac{2Q}{2m-1 +Q}\right)^\alpha\right] \, .
\label{eq:fhmf1}
\end{equation}
Eq.\,\eqref{eq:fhmf1} can also have either positive and negative values. Let $\tilde{\alpha}(Q)$ the value of $\alpha$ for which the right-hand side of Eq.\,\eqref{eq:fhmf1} vanishes for a fixed $Q$, then
\begin{equation}
\label{eq:tildealpha}
\tilde{\alpha}(Q) = \frac{\log\left(1-\frac{1}{m}\right)}{\log\left(1-\frac{2Q}{2m-1+ Q}\right)}\, .
\end{equation}
We have $\tilde{\alpha}(1)=1$ and $\tilde{\alpha}(Q)\rightarrow \infty$ if $Q\rightarrow 0^+$. The function $\tilde{\alpha}(Q)$ is drawn in black in Figs.~\ref{fig:stability01mreg}a-c and it delimits the red region where $(f_\alpha^{(hmf)})^\prime(1)<0$ ($\hat{a}^*_{2m}=1$ is stable) and the blue region where $(f_\alpha^{(hmf)})^\prime(1)>0$ ($\hat{a}^*_{2m}=1$ is unstable).


\subsubsection*{Generalise to complete graphs}

Let us conclude this part by showing the previous analysis returns the results obtained by using the mean-field hypothesis once we assume the underlying network to be a complete graph. To simplify the setting we will assume the network to be composed of $N=2N'+1$ nodes and $m=N'$, hence each node has $2N'$ neighbors. Let us observe that one trivially has $p_{k}=1$ if $k=N-1=2N'$ and $p_k=0$ otherwise, and $\langle k\rangle = N-1=2N'$, which implies that $q_k=1$ if $k=N-2=2N'-2=2(m-1)$ and $0$ otherwise. From Eq.~\eqref{eq:amean} we can obtain $\langle a\rangle = a_{N-1}=n_A/N$, namely there is only one variable that is the proportion of agents with opinion $A$. From the Eq.~\eqref{eq:fsimp} we can get 
\begin{equation*}
f_\alpha^{(hmf)}(a_{N-1})= -a_{N-1}+\sum_{\omega=0}^{2N'} \binom{2N'}{\omega} (a_{N-1})^{2N'-\omega}\left(1-a_{N-1}\right)^\omega P_\alpha\left( \frac{n_A}{n_A +n_B Q}\right)\, ,
\end{equation*}
where we recall that $\omega=n_B$, to be the number of agents with opinion $B$, and $2N'-\omega=n_A$, the number of agents with opinion $A$. Let us observe that we also added the term $\omega=2N'$ in the sum, whose contribution vanishes because $P_\alpha$ does. By using $n_A+n_B=N$ we can rewrite the previous equation as
\begin{eqnarray*}
f_\alpha^{(hmf)}(a_{N-1})&=& -a_{N-1}+\sum_{\omega=0}^{2N'} \binom{2N'}{\omega} (a_{N-1})^{2N'-\omega}\left(1-a_{N-1}\right)^\omega P_\alpha\left( \frac{n_A}{n_A (1-Q)+N Q}\right)\\
&=& -a_{N-1}+\sum_{\omega=0}^{2N'} \binom{2N'}{\omega} (a_{N-1})^{2N'-\omega}\left(1-a_{N-1}\right)^\omega   P_\alpha\left( \frac{a_{N-1}}{a_{N-1}(1-Q)+Q}\right)\, ,
\end{eqnarray*}
where in the last step we divided by $N$ the number of agents to obtain the proportion. Being the term involving $P_\alpha$ independent from $\omega$, we eventually obtain
\begin{equation*}
\begin{split}
f_\alpha^{(hmf)}(a_{N-1}) &=-a_{N-1}+P_\alpha\left( \frac{a_{N-1}}{a_{N-1}(1-Q)+Q}\right)\sum_{\omega=0}^{2N'} \binom{2N'}{\omega} (a_{N-1})^{2N'-\omega}\left(1-a_{N-1}\right)^\omega=\\ &=-a_{N-1}+P_\alpha\left( \frac{a_{N-1}}{a_{N-1}(1-Q)+Q}\right)\, ,
\end{split}
\end{equation*}
where the Newton property for the binomial has been used to eventually get the same function we obtained under the mean field assumption~\eqref{eq:odea2}.

Let us now rewrite $\hat{\alpha}(Q)$ given by Eq.~\eqref{eq:hatalpha} under the above assumption of complete graph, namely
\begin{equation*}
\hat{\alpha}(Q) = \frac{\log\left(1-\frac{1}{N'}\right)}{\log\left(1-\frac{2}{1+(2N'-1) Q}\right)}\, ,
\end{equation*}
then letting the number of nodes to be very large, $N=2N'+1\rightarrow \infty$, then we obtain
\begin{equation*}
\hat{\alpha}(Q) \sim Q+\dots \, ,
\end{equation*}
namely the curve separating the convergence to a consensus for option $A$ (red region in Fig.~\ref{fig:stability01mreg}) to the region where mistakes are possible (green region in Fig.~\ref{fig:stability01mreg}) converges to the line $\hat{\alpha}(Q) = Q$ in the limit of infinitely many agents, in agreement with the results reported in Fig.~\ref{fig:stability01complete}.

{\color{blue}
\section{Degree preserving rewiring process}
This section briefly presents the degree preserving rewiring process, sometimes called criss-cross in the literature, used in the main text to reduce the average shortest path while keeping the degree distribution unchanged.

As shown in Fig.\,\ref{fig:crisscross}, let us consider four nodes, $i_1,i_2,i_3$ and $i_4$ such that $i_1$ and $i_2$ are connected to each other by an edge but not to $i_3$ and $i_4$, and vice-versa, nodes $i_3$ and $i_4$ are connected to each other by an edge but not to $i_1$ and $i_2$. Then, in the criss-cross process, the two existing edges are deleted and two new edges are added from $i_1$ to $i_4$ and from $i_2$ to $i_3$. In such a way the four nodes will not change their degree and therefore $p_k$ also remains unchanged.
\begin{figure*}[ht!]
    \centering
    \includegraphics[width=0.5\textwidth]{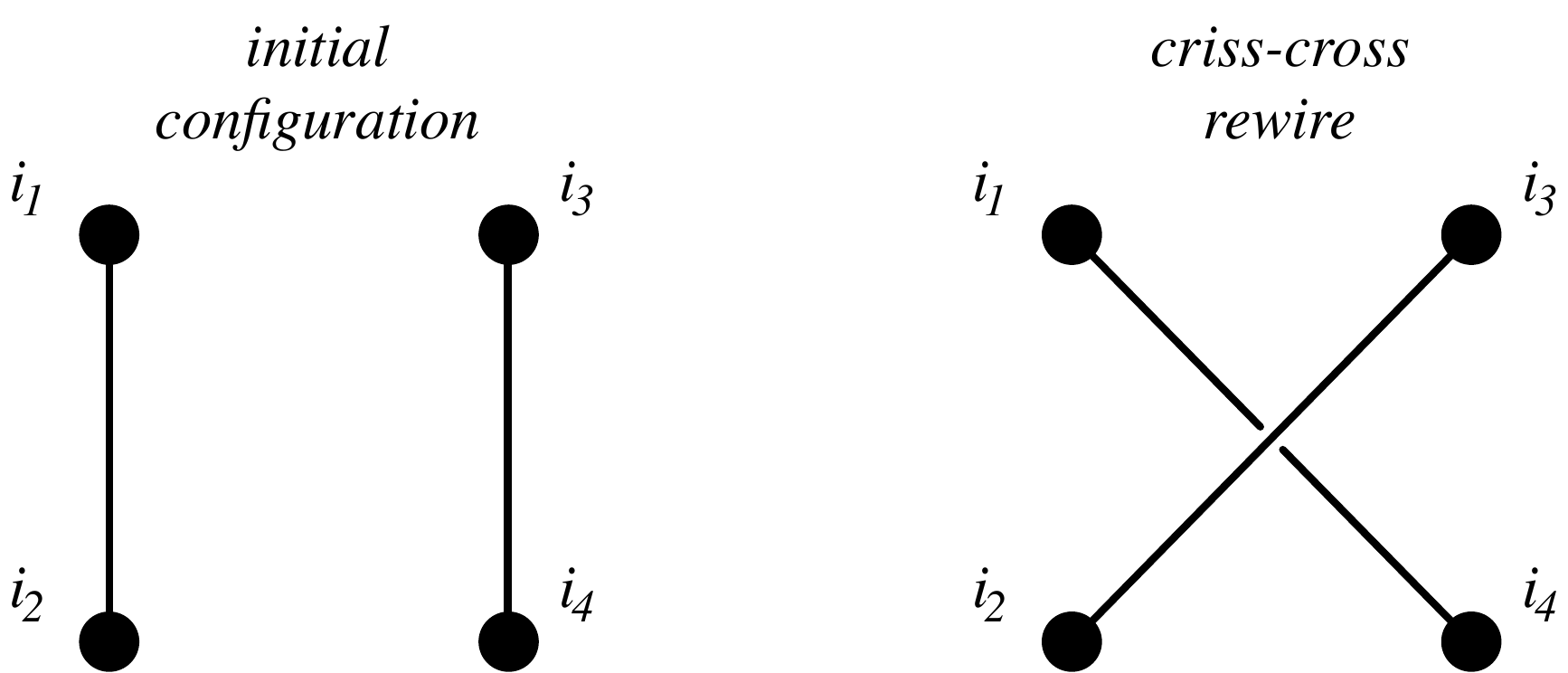}
    \caption{The criss-cross rewiring process. The initial configuration (left panel) has the edges $(i_1,i_2)$ and $(i_3,i_4)$, after the rewiring process (right panel) those edges are deleted and replaced with the new ones $(i_1,i_4)$ and $(i_2,i_3)$.}
        \label{fig:crisscross}
\end{figure*}
}

{\color{blue}
\section{Scale-free networks}

Sec.\,\ref{ssec:scale-free} in the main text presents our analysis of the population dynamics on scale-free networks. While Figs.\,\ref{fig:stability01scalefreeGamma} and~\ref{fig:stability01scalefreeQ} show the results for scale-free network with minimum degree $k_{min}=2$, here we complement the analysis providing the results for $k_{min}=1$ and $k_{min}=3$. The multiagent simulations for $k_{min}=2$ and $k_{min}=3$ have been performed on scale-free networks built using the configuration model~\cite{Newman}, instead, as also indicated in the main text, for $k_{min}=1$ the scale-free networks have been built using the Simon model~\cite{simon1955,Newman}. 

For low values of $k_{min}$, i.e., for $k_{min}\le2$, it is possible to appreciate an important change in the dynamics for different values of the connectivity parameter $\gamma$. Instead, for $k_{min}=3$, the parameter $\gamma$ has a much smaller effect on the population dynamics, as clearly shown by almost-horizontal separation lines between the colored areas in Fig.\,\ref{fig:scalefreeQ-kmin3}.

\begin{figure*}[ht!]
    \centering
    \includegraphics[width=\textwidth]{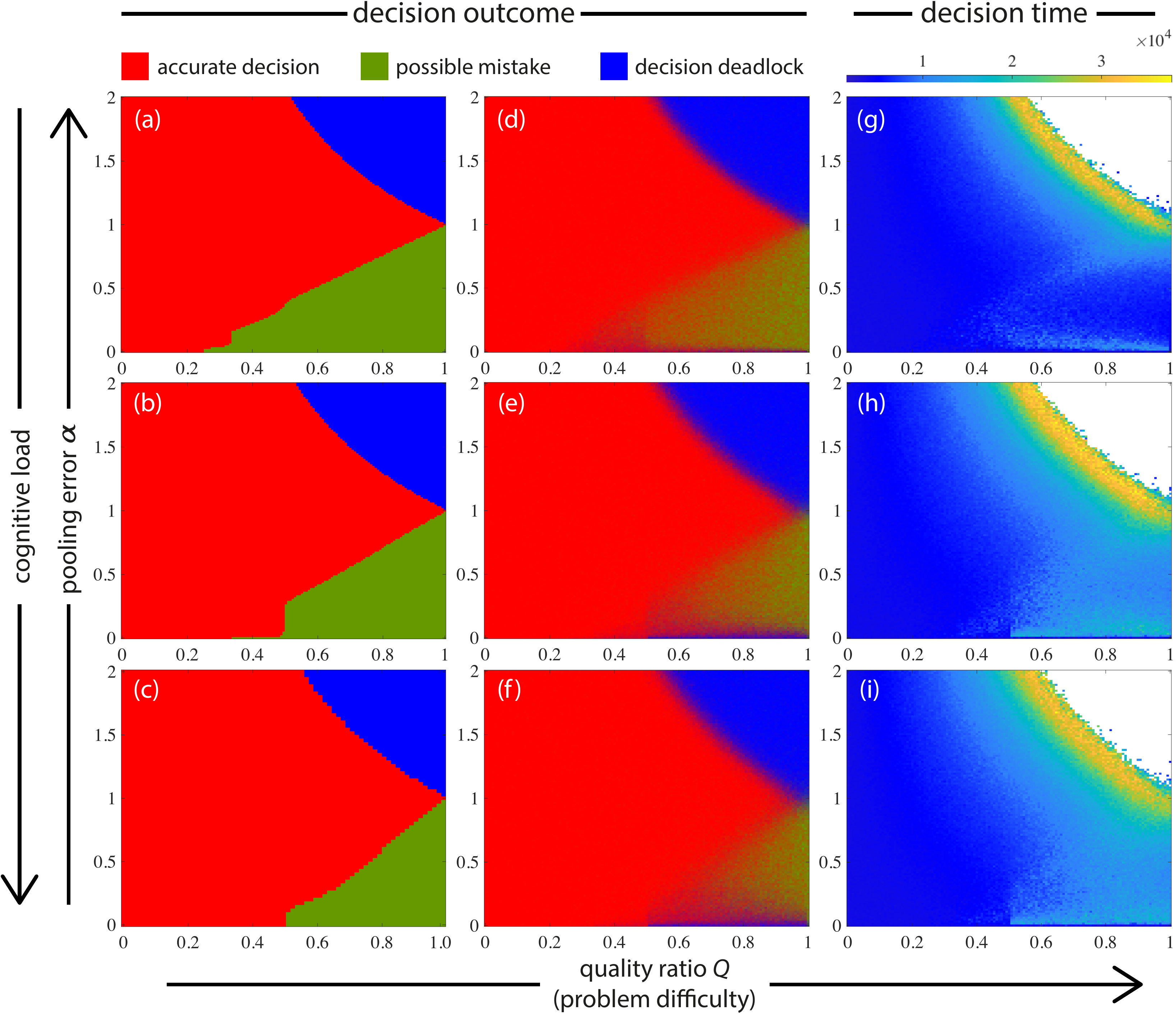}
        \caption{Stability diagrams (panels a-c), decision outcome from multiagent simulations (panels d-f) and convergence time (panels g-i) for collective decision-making on scale-free networks with $k_{min}=1$ as a function of the pooling error $\alpha$ and quality ratio $Q$. We report results for three values of the exponent $\gamma$ regulating network connectivity: top row $\gamma=2.2$, central row $\gamma=2.6$, bottom row $\gamma=3.1$. Color code and experimental design are the same as the one described in the caption of Fig.\,\ref{fig:stability01scalefreeGamma}.} 
        \label{fig:scalefreeGamma-kmin1}
\end{figure*}

\begin{figure*}[ht!]
    \centering
    \includegraphics[width=\textwidth]{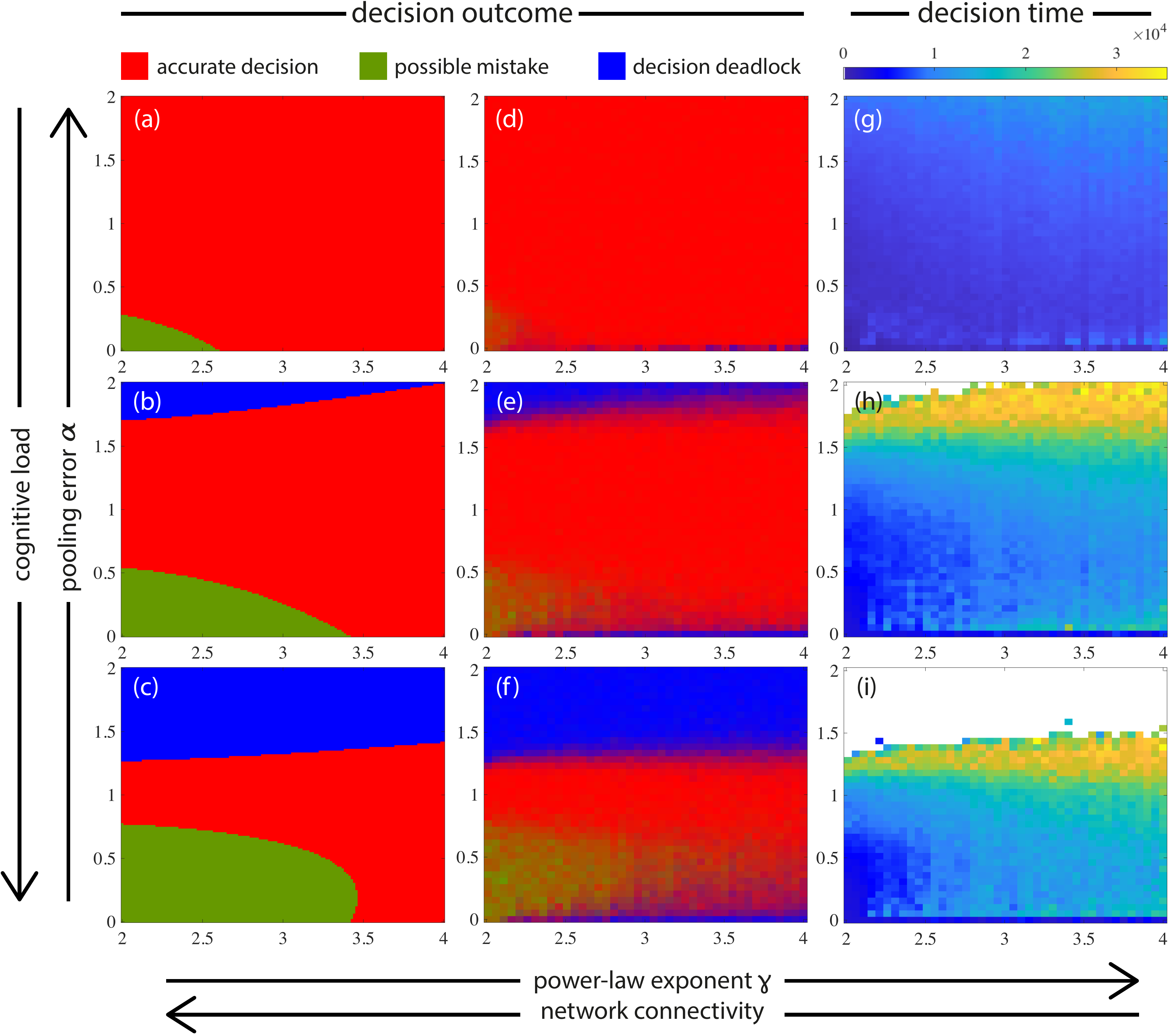}
    \caption{Stability diagrams (panels a-c), decision outcome from multiagent simulations (panels d-f) and convergence time (panels g-i) for collective decision-making on scale-free networks with $k_{min}=1$ as a function of the pooling error $\alpha$ and network's power-law exponent $\gamma$. We report results for three values of the quality ratio $Q=Q_B/Q_A$ which encodes the decision problem difficulty: top row $Q=0.5$ (easy problem), central row $Q=0.8$ (medium problem), bottom row $Q=0.9$ (difficult problem), with $Q_A=1$. Color code and experimental design are the same as the one described in the caption of Fig.\,\ref{fig:stability01scalefreeGamma}.}
    \label{fig:scalefreeQ-kmin1}
\end{figure*}

\begin{figure*}[ht!]
    \centering
    \includegraphics[width=\textwidth]{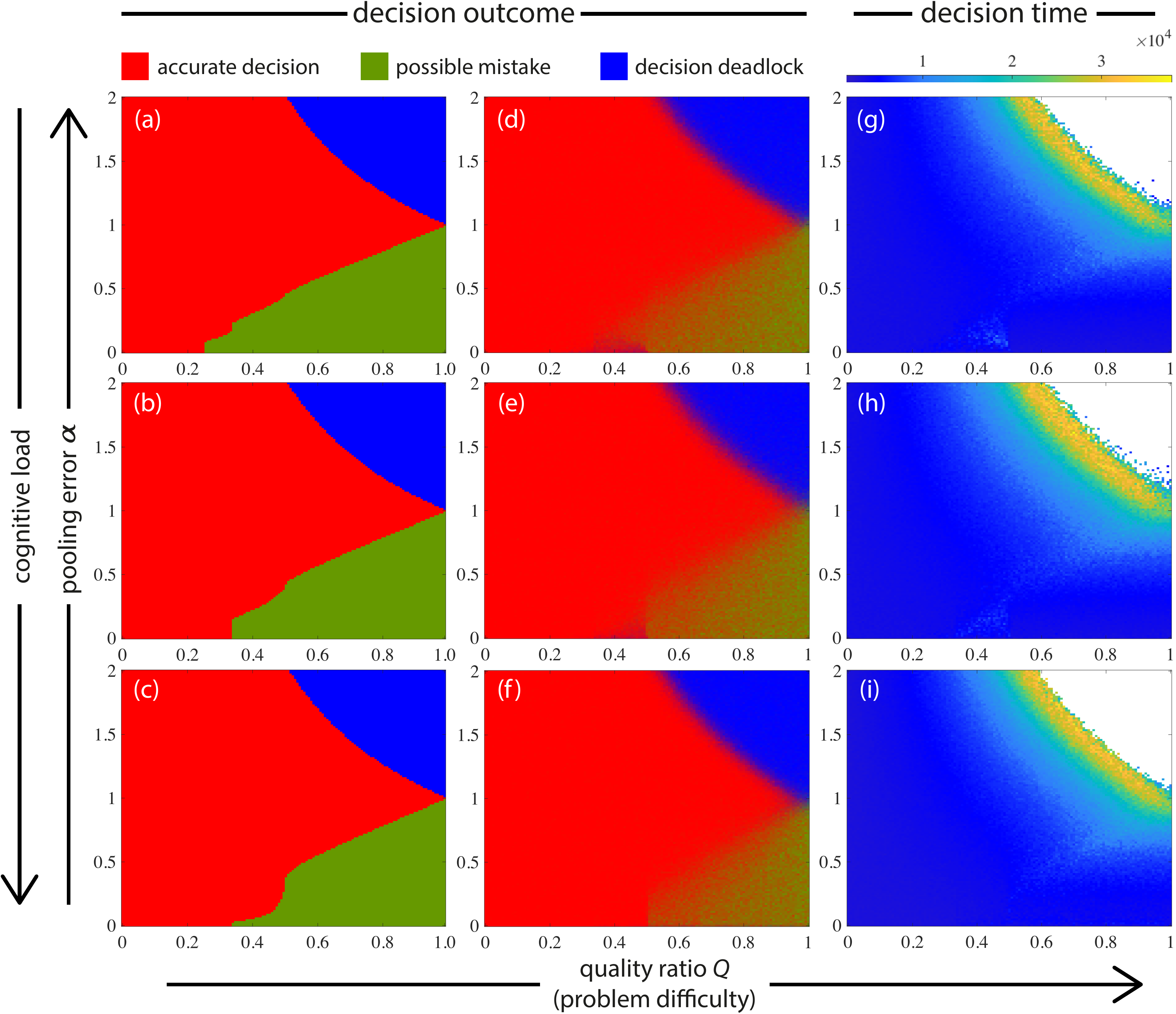}
        \caption{Stability diagrams (panels a-c), decision outcome from multiagent simulations (panels d-f) and convergence time (panels g-i) for collective decision-making on scale-free networks with $k_{min}=3$ as a function of the pooling error $\alpha$ and quality ratio $Q$. We report results for three values of the exponent $\gamma$ regulating network connectivity: top row $\gamma=2.2$, central row $\gamma=2.6$, bottom row $\gamma=3.1$. Color code and experimental design are the same as the one described in the caption of Fig.\,\ref{fig:stability01scalefreeGamma}.} 
        \label{fig:scalefreeGamma-kmin3}
\end{figure*}

\begin{figure*}[ht!]
    \centering
    \includegraphics[width=\textwidth]{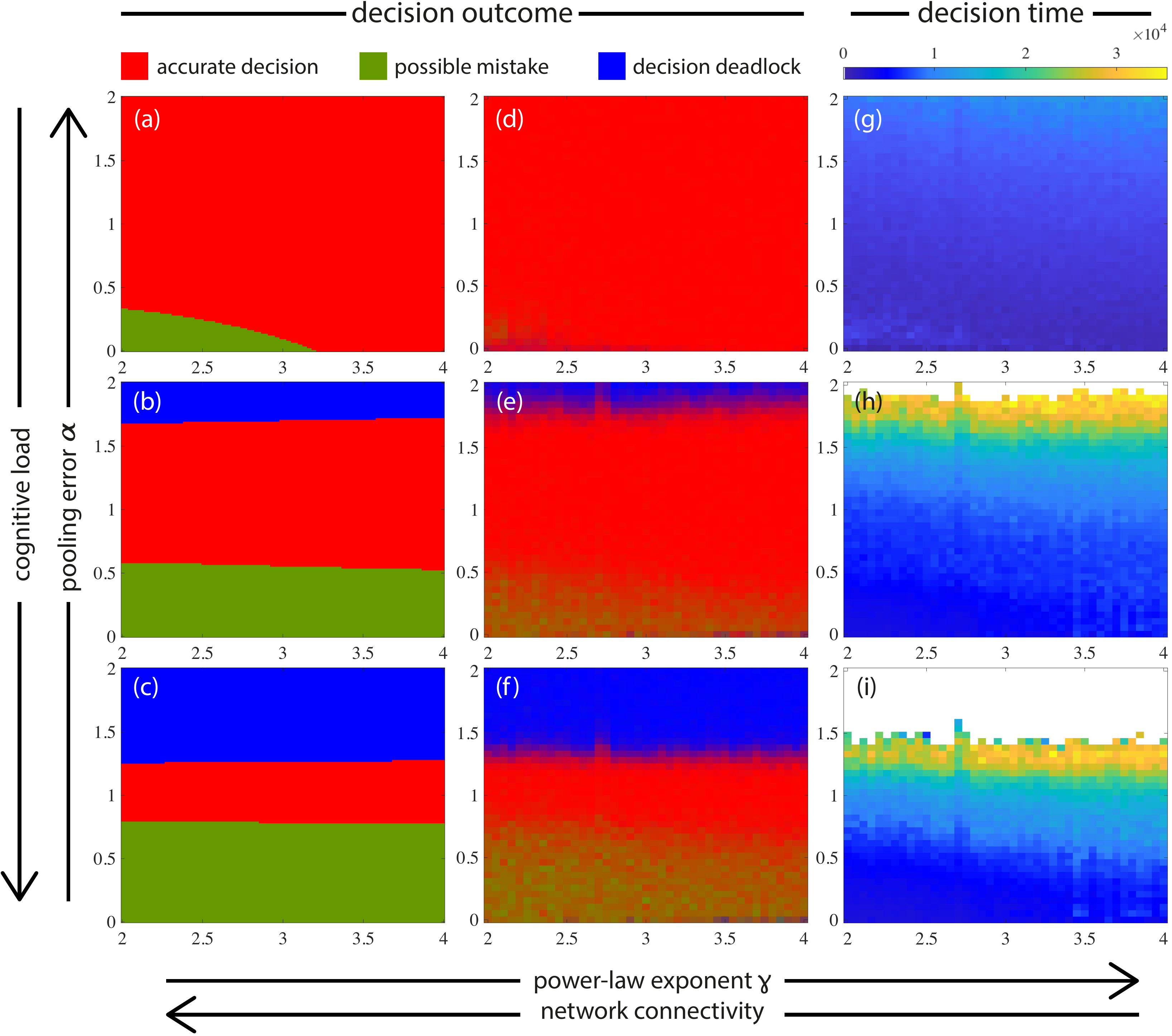}
    \caption{Stability diagrams (panels a-c), decision outcome from multiagent simulations (panels d-f) and convergence time (panels g-i) for collective decision-making on scale-free networks with $k_{min}=3$ as a function of the pooling error $\alpha$ and network's power-law exponent $\gamma$. We report results for three values of the quality ratio $Q=Q_B/Q_A$ which encodes the decision problem difficulty: top row $Q=0.5$ (easy problem), central row $Q=0.8$ (medium problem), bottom row $Q=0.9$ (difficult problem), with $Q_A=1$. Color code and experimental design are the same as the one described in the caption of Fig.\,\ref{fig:stability01scalefreeGamma}.}
    \label{fig:scalefreeQ-kmin3}
\end{figure*}
}

{\color{blue}
\section{Erd\H{o}s-R\'enyi random graph}
\label{sec:ERgraph}
The aim of this section is to perform an analysis similar to the one presented in the main text in the case of scale-free networks and $2m$-regular graphs, but by assuming agents to interact via an Erd\H{o}s-R\'{e}nyi random graph, composed of $N$ nodes and where each couple of nodes has a probability $p>0$ to be connected. In Fig.\,\ref{fig:ER} we report the results obtained by assuming the HMF hypothesis (panels (a) and (d)) and we compare them with the multiagent numerical simulations performed on an Erd\H{o}s-R\'{e}nyi random graph comprising $N=200$ nodes and $p=0.02$ (panel (b)) and $p=0.2$ (panel (e)). We can observe that the agreement is good, especially in the case of $p=0.2$, which corresponds to a network with a larger average degree, $\langle k\rangle \sim 40$, and thus to a smaller averaged shortest path (compare also with panel (a) of Fig.\,\ref{fig:stability01complete} in the main text). In panels (c) and (f) we report the converge time to the consensus states, all-A or all-B, and we can again observe that approaching the deadlock decision regions, the convergence time increases.
\begin{figure*}[ht!]
    \centering
    \includegraphics[width=0.8\textwidth]{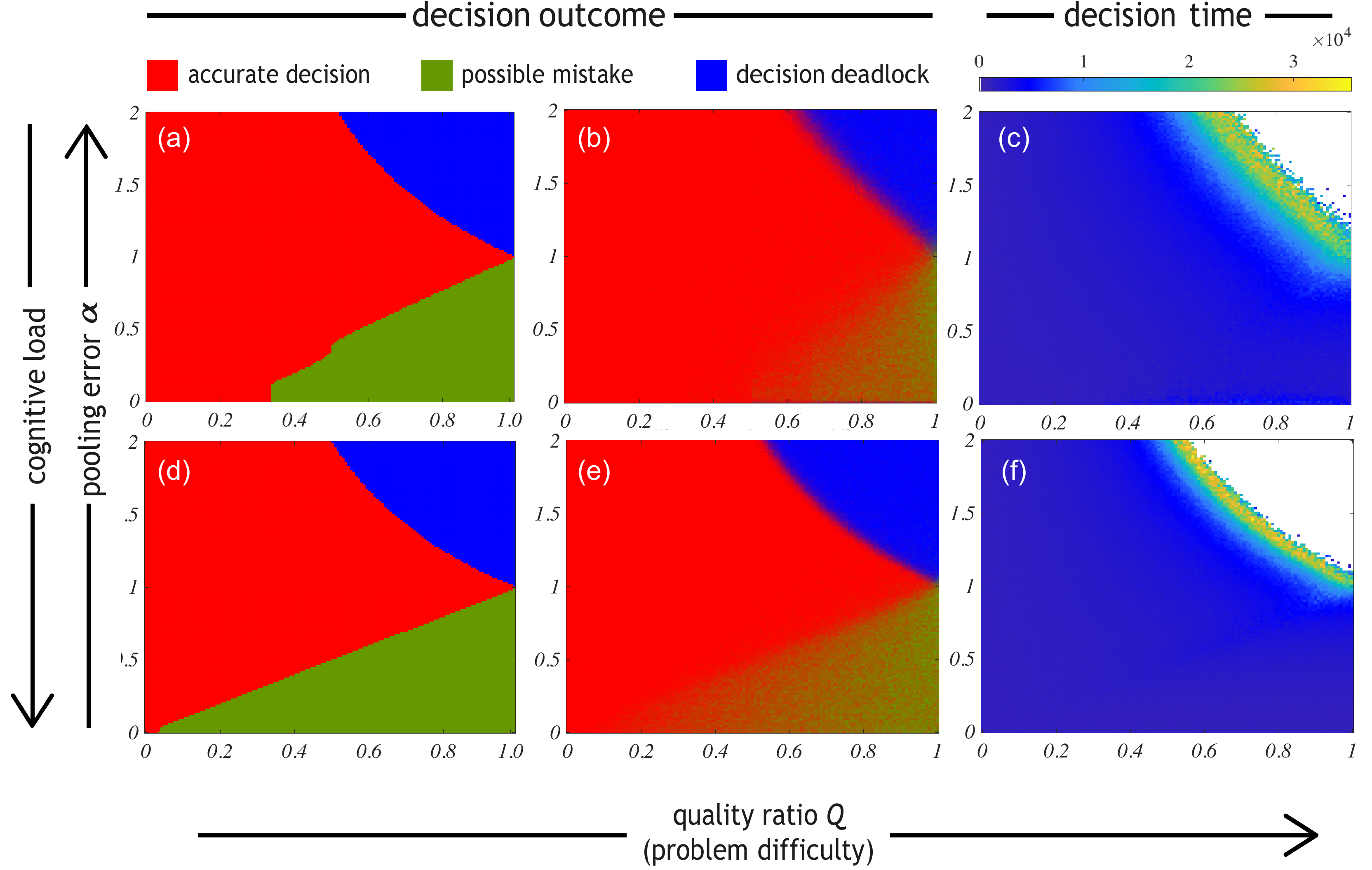}
    \caption{Stability diagrams (panels a,d), decision outcome from multiagent simulations (panels b,e) and convergence time (panels c,f) for collective decision-making on Erd\H{o}s-R\'{e}nyi random graph as a function of the pooling error $\alpha$ and quality ratio $Q$. We present the results for two values of the probability $p$ regulating network connectivity: top row $p=0.02$ and bottom row $p=0.2$, the number of nodes has been fixed to $N=200$. Left column panels -- i.e., (a) and (d) -- show the convergence diagram of the mean-field model~\eqref{eq:dakdt3b}. The parameter space is divided into the same three regions of Fig.\,\ref{fig:stability01complete}a using the same color code. Central column panels show the results of simulations ($100$ independent for each $(Q,\alpha)$ configuration) of $N=200$ agents interacting on a Erd\H{o}s-R\'{e}nyi random graph with random initial configurations (i.e., $n_A(t=0) \sim \mathcal{U}(0,N)$) for $50\,000$ time steps. Central column panels -- i.e., (b) and (e) -- show the outcome of the collective decision-making process using the same RGB color code as Fig.\,\ref{fig:stability01complete}d. Right column panels -- i.e., (c) and (f) -- show the average number of timesteps needed to reach a consensus, i.e., $n_A=200$ or $n_B=200$. The white area indicates the absence of data, as the system never reaches a consensus.}
        \label{fig:ER}
\end{figure*}

Fig.\,\ref{fig:ERp} shows the combined impact of the probability to have an edge, $p$, and the pooling error $\alpha$ on the system outcome once we fix $Q_B$ and $Q_A$. Let us observe that for an easy problem ($Q_B=0.2$ top row), agents are almost always able to reach a consensus for the better quality option (red region), only for very small $\alpha$ mistakes are possible, moreover those behaviors do not seem to depend on $p$. On the other hand for an hard task ($Q_B=0.8$ bottom row), the diagram is divided into three zones and only for $\alpha$ close to $1$, namely once agents adopt the majority rule, they are able to converge to the best option. For smaller $\alpha$ mistakes are possible, while for larger $\alpha$ a deadlock outcome is obtained. Again those results seem not to depend on $p$.
\begin{figure*}[ht!]
    \centering
    \includegraphics[width=0.8\textwidth]{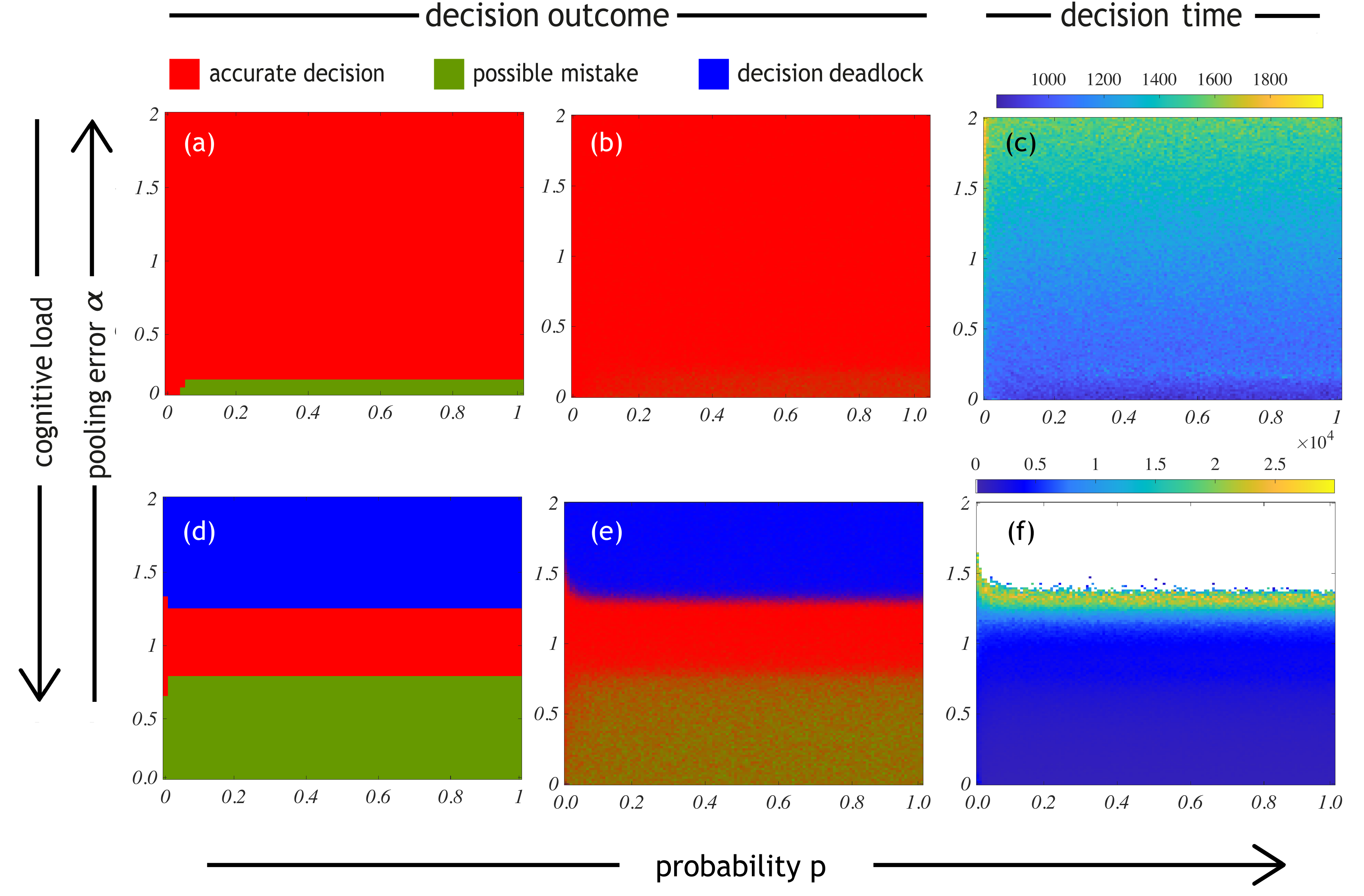}
    \caption{Stability diagrams (panels a,d), decision outcome from multiagent simulations (panels b,e) and convergence time (panels c,f) for collective decision-making on Erd\H{o}s-R\'{e}nyi random graph as a function of the pooling error $\alpha$ and the probability $p$ to establish a link between two nodes. We present the results for two values of the quality option $Q_B$ for fixed $Q_A=1$: top row $Q_B=0.2$ and bottom row $Q_B=0.8$, the number of nodes has been fixed to $N=200$. Left column panels -- i.e., (a) and (d) -- show the convergence diagram of the mean-field model~\eqref{eq:dakdt3b}. The parameter space is divided into the same three regions of Fig.\,\ref{fig:stability01complete}a using the same color code. Central column panels show the results of simulations ($100$ independent for each $(p,\alpha)$ configuration) of $N=200$ agents interacting on a Erd\H{o}s-R\'{e}nyi random graph with random initial configurations (i.e., $n_A(t=0) \sim \mathcal{U}(0,N)$) for $50\,000$ time steps. Central column panels -- i.e., (b) and (e) -- show the outcome of the collective decision-making process using the same RGB color code as Fig.\,\ref{fig:stability01complete}d. Right column panels -- i.e., (c) and (f) -- show the average number of timesteps needed to reach a consensus, i.e., $n_A=200$ or $n_B=200$. The white area indicates the absence of data, as the system never reaches a consensus.}
    \label{fig:ERp}
\end{figure*}
}

\end{document}